\documentclass[journal=jctcce, manuscript=article, layout=twocolumn, email=true]{achemso}
\setkeys{acs}{articletitle=true}
\usepackage{amsfonts,amssymb,amsmath}
\usepackage[english]{babel}
\usepackage[utf8x]{inputenc}
\usepackage[T1]{fontenc}

\usepackage{threeparttable}
\usepackage{lmodern}
\usepackage{dcolumn}
\usepackage{graphicx}
\usepackage[version=3]{mhchem}
\usepackage[normalem]{ulem}
\usepackage{paralist}
\usepackage{lineno,xcolor}
\usepackage{cancel}
\usepackage{siunitx}
\usepackage{float}
\usepackage[colorlinks,
 allcolors=blue,
 citecolor=blue,
 urlcolor=blue,
 bookmarks]{hyperref}

\usepackage{booktabs}
\usepackage{caption,setspace}
\usepackage{tabularx}
\usepackage{multirow}
\usepackage{bm}

\usepackage{placeins}

\DeclareMathOperator{\Tr}{Tr}

\newcommand{\Matrix}[1]{\ensuremath{\bm{#1}}}

\title{Random-Phase Approximation in Many-Body Noncovalent Systems: Methane in a Dodecahedral Water Cage}

\author{Marcin Modrzejewski}
\email{m.m.modrzejewski@gmail.com}
\affiliation{Faculty of Chemistry, University of Warsaw, 02-093 Warsaw, Pasteura 1, Poland}
\alsoaffiliation{Department of Chemical Physics and Optics, Faculty of Mathematics and Physics, Charles University, Ke Karlovu 3, CZ-12116
Prague 2, Czech Republic}
\author{Sirous Yourdkhani}
\email{yourdkhani.sirous@karlov.mff.cuni.cz}
\affiliation{Department of Chemical Physics and Optics, Faculty of Mathematics and Physics, Charles University, Ke Karlovu 3, CZ-12116
Prague 2, Czech Republic}
\author{Szymon Śmiga}
\affiliation{Institute of Physics, Faculty of Physics, Astronomy and Informatics,
Nicolaus Copernicus University, Grudziądzka 5, 87-100 Toruń, Poland}
\author{Jiří Klimeš}
\email{klimes@karlov.mff.cuni.cz}
\affiliation{Department of Chemical Physics and Optics, Faculty of Mathematics and Physics, Charles University, Ke Karlovu 3, CZ-12116
Prague 2, Czech Republic}

\begin{document}
\begin{abstract}
  The many-body expansion (MBE) of energies of molecular clusters or solids
offers a way to detect and analyze errors of theoretical methods that 
could go unnoticed if only the total energy of the system was considered.
In this regard, the interaction between the methane molecule and its enclosing dodecahedral water
cage, \ce{CH4\bond{...}(H2O)20}, is a stringent test for approximate methods, including 
density-functional theory (DFT) approximations.
Hybrid and semilocal DFT approximations behave erratically for this system,
with three- and four-body nonadditive terms having neither the correct sign
nor magnitude.
Here we analyze to what extent these qualitative errors in different MBE contributions are conveyed
to post-Kohn-Sham random-phase approximation (RPA), which uses approximate
Kohn-Sham orbitals as its input.
The results reveal a correlation between the quality 
of the DFT input states and the RPA results.
Moreover, the renormalized singles energy (RSE) corrections 
play a crucial role in all orders of the many-body expansion.
For dimers, RSE corrects the RPA underbinding for every tested
Kohn-Sham model: generalized-gradient approximation (GGA), meta-GGA, (meta-)GGA hybrids, as well as
the optimized effective potential at the correlated level.
Remarkably, the inclusion of singles in RPA can also correct
the wrong signs of three- and four-body nonadditive energies as well as
mitigate the excessive higher-order contributions to
the many-body expansion.
The RPA errors are dominated by the contributions of compact clusters.
As a workable method for large systems, we propose to replace those
compact contributions with CCSD(T) energies and to sum up the remaining
many-body contributions up to infinity with supermolecular or periodic RPA.
As a demonstration of this approach, we show that for RPA(PBE0)+RSE
it suffices to apply CCSD(T) to dimers and 30 compact, hydrogen-bonded
trimers to get the methane--water cage interaction energy 
to within 1.6~\% of the reference value.

\end{abstract}

\maketitle

\section{Introduction}

%
Datasets of interaction energies of molecular dimers are widely used to assess the accuracy of quantum chemistry 
methods for noncovalent interactions.\cite{jurecka2006benchmark,peverati2014quest,taylor2016blind,goerigk2017a}
How the methods perform for nonadditive interactions, that is, interactions involving three or more molecules
with pairwise contributions removed, has gained much less attention.\cite{rezac2015benchmark,huang2015reliable,jankiewicz2018dispersion}
However, it was shown that three-body nonadditive interactions make an important contribution 
to the binding energies of atomic and molecular solids.\cite{lotrich1997three,yang2014determination,kennedy2014resolving}
Therefore, a reliable description of many-body interactions is a clear aim in the development
of low-scaling methods for large clusters and molecular solids.\cite{rezac2015benchmark}
The three-body nonadditive energies pose a challenge to density functional theory (DFT)
approximations.
While it is widely known that approximate DFT functionals lack the many-body dispersion energy,
an even larger shortcoming lies in the spurious many-body contribution from the semilocal exchange
energy approximation.\cite{gillan2014many,hapka2017nature,jankiewicz2018dispersion}
One way to improve the results is to use a scheme that does not approximate electron exchange,
such as the random-phase approximation (RPA).
RPA is based on the frequency-dependent
density response function built from Kohn-Sham orbitals and orbital energies.\cite{eshuis2012electron}
The nonlocal RPA correlation energy is compatible with the exact Kohn-Sham exchange energy.\cite{dobson2012dispersion}
Due to the account of $n$-body dispersion,\cite{dobson2014beyond,hermann2017first} RPA's description of molecular
solids\cite{huang2015reliable} and large clusters\cite{richard2014aiming}
is more adequate than that of commonly used second-order M{\o}ller-Plesset perturbation theory (MP2).
At the same time, low-scaling implementations enable routine use of RPA for molecular
clusters and solids.\cite{eshuis2010fast,delben2013electron,kaltak2014low,kaltak2014cubic,schurkus2016effective,wilhelm2016large,klimes2016lattice,modrzejewski2020random} 
It needs to be added that the coarse-grained approximation to the RPA energy
has been the formal foundation of the forcefield-like many-body dispersion
corrections.\cite{ambrosetti2014long,hermann2017first,kim2020umbd}

In our recent work,\cite{modrzejewski2020random} we have shown that RPA is highly
reliable for hydrogen-bonded trimers, where many-body effects
are mostly due to polarization.\cite{hesselmann2018correlation}
However, the systems with medium- and high-dispersion content
prove to be more difficult.\cite{modrzejewski2020random}

Methane clathrate, modeled either as a finite cluster\cite{deible2014theoretical,gillan2015energy} or as a bulk 
material,\cite{cox2014benchmarking} is a particularly difficult case for widely
used theoretical methods.
For bulk clathrate, Cox~{\it et al.}\cite{cox2014benchmarking} found that none of the tested DFT methods, including
nonlocal van der Waals functionals, correctly describes both the methane binding energy
and lattice constants of the clathrate.
The lack of a reliable description of nonadditive interactions is one
of the main causes of those errors.
As shown by Deible~{\it et al.}\cite{deible2014theoretical} for a finite cluster, \ce{CH4\bond{...}(H2O)20},
the semilocal DFT approximation breaks down for the three-body contribution of the methane binding energy,
where most models predict a wrong sign and/or order of magnitude.
Importantly, the major part of those errors was shown to originate from the approximations
in the exchange functional, which cannot be simply corrected by an addition of
one of available many-body dispersion corrections.\cite{hermann2017first}

In this work, we report a new CCSD(T) estimate of the many-body expansion contributions
to the methane binding energy in the \ce{CH4\bond{...}(H2O)20} cluster.
Subsequently, we use the reference coupled-clusters data to determine how the performance of post-Kohn-Sham RPA
depends on the approximate orbitals and orbital energies employed as its input.
The assessment comprises the eigenstates generated with the PBE,\cite{perdew1996generalized} PBE0,\cite{adamo1999toward}
SCAN,\cite{sun2015strongly} and SCAN0\cite{hui2016scan}
exchange-correlation models, which covers the generalized-gradient, meta-generalized-gradient,
and hybrid rungs of the DFT ladder of approximations.\cite{perdew2005prescription}
Special focus is on the effect of the renormalized singles energy (RSE) correction
in many-body noncovalent energy contributions.\cite{klimes2015singles,ren2013renormalized,modrzejewski2020random}
Finally, we examine the performance of the incremental correction scheme, where the low-order
many-body contributions at a higher level of theory are combined with the remaining
contributions estimated from a supermolecular RPA calculation.
The incremental approach potentially decreases the number of fragment contributions that need to be 
obtained explicitly.
This reduces the computational time and also decreases the numerical noise originating from 
summation of a large number of small contributions.

\section{Methods}
We consider the interaction energy $E_\text{int}$ between the methane molecule
and the \ce{(H2O)20} cage:
\begin{equation}
E_\text{int}= E_{\rm CH_4\ce{(H2O)20}}-E_{\rm CH_4}-E_{\rm \ce{(H2O)20}}
\label{eq:supermolecular}
\end{equation}
where $E_{\rm CH_4\ce{(H2O)20}}$ is the energy of the methane in the water cage,
$E_{\rm CH_4}$ and $E_{\rm \ce{(H2O)20}}$ are the energies of isolated methane
molecule and water cage, respectively.
%
Within the many-body expansion approach the methane molecule in 
a dodecahedral water cage is treated as a cluster of $N=21$ molecules.
The interaction energy between methane and the water cage is assembled from $n$-body contributions, 
where each $n$-body fragment includes methane:
\begin{align}
 E_\text{int} &= E_\text{int}[2] + E_\text{int}[3] + E_\text{int}[4] + \ldots \\
 E_\text{int}[2] &= \sum_{i=1}^{20} E_\text{int}(0, i) \\
 E_\text{int}[3] &= \sum_{i=1}^{19} \sum_{j=i+1}^{20} E_\text{int}(0, i, j) \label{trimer-energies} \\
 E_\text{int}[4] &= \sum_{i=1}^{18} \sum_{j=i+1}^{19} \sum_{k=j+1}^{20} E_\text{int}(0, i, j, k) \label{tetramer-energies}
\end{align}
The indices $i$, $j$, $k$, {\ldots} denote the individual water molecules;
0 corresponds to the methane molecule.
The brackets are dropped from the notation when no ambiguity arises.
$E_\text{int}(0, i)$ denotes the dimer interaction energy of methane and $i$th water
\begin{equation}
E_\text{int}(0, i) = E(0, i) - E(0) - E(i) \label{two-body-energy}
\end{equation}
The higher-order terms are the nonadditive interaction energies
of individual trimers
\begin{multline}
 E_\text{int}(0, i, j) = E(0, i, j)\\
       - E(0) - E(i) - E(j) \\
      - E_\text{int}(0, i) - E_\text{int}(0, j) - E_\text{int}(i, j) \label{three-body-energy}
\end{multline}
and tetramers
\begin{multline}
 E_\text{int}(0, i, j, k) = E(0, i, j, k) \\
     - E(0) - E(i) - E(j) - E(k) \\
   - E_\text{int}(0, i)  - E_\text{int}(0, j) - E_\text{int}(0, k) \\
    - E_\text{int}(i, j)  - E_\text{int}(i, k) - E_\text{int}(j, k) \\
   - E_\text{int}(0, i, j) - E_\text{int}(0, i, k) \\
   - E_\text{int}(0, j, k) - E_\text{int}(i, j, k) \label{four-body-energy}
\end{multline}
The energy of geometry relaxation is not included, {\it i.e.}, the coordinates are fixed
at their cluster values.
All energies contributing to an $n$-body term are computed in the basis
set of the corresponding $n$-body cluster.
We compare the different choices of basis sets below in Sec.~\ref{sec:bsse}.
The explicit many-body expansion includes all clusters up to tetramers: 20 dimers, 190 trimers, 
and 1140 tetramers.
%

In the text and tables, we refer to the set of all $n$-body
clusters using the label $n$b.
Furthermore, the clusters with more than one water molecule
are divided into subsets according to the number of hydrogen bonds.
The label of an $n$-body cluster with $m$ hydrogen
bonds is $n$b $m$hb.
For example, the label corresponding to the tetramers \ce{CH4(H2O)3}
having two hydrogen bonds is 4b~2hb.

For the Hartree-Fock, MP2, DFT, and RPA methods it is computationally
possible to obtain the energy of the complete cluster.
In such a case the interaction energy is obtained directly by applying Eq.~\ref{eq:supermolecular}.
As the water cage is treated as a single fragment, this scheme is referred to as the 
supermolecular approach.
The supermolecular interaction energy is also used to provide an estimate of five-
and higher-body terms. 
The basis set used to perform the calculations needed to evaluate the supermolecular 
interaction energy includes basis functions of all the atoms of the complex (full cluster basis set).

The Hartree-Fock, M{\o}ller-Plesset, and coupled cluster calculations were done
using Molpro.\cite{werner2012molpro}
We used the augmented version of the correlation-consistent basis sets of
Dunning and coworkers,\cite{kendall1992electron,schuchardt2007basis}
which we denote by a shorthand AV$n$Z, where $n$ is the cardinal number of the basis set.
Only the valence electrons were considered to obtain the correlation energy.
The correlation energy was extrapolated to the basis set limit using the 
two-point formula of Halkier {\it et al.} assuming that the basis-set error is proportional to $n^{-3}$.\cite{halkier1999basis}
As an alternative, the explicitly correlated (F12) approach was used to speed-up the basis-set 
convergence of the coupled cluster and MP2 energies.\cite{werner2007general,adler2007simple,knizia2009simplified}
For the mean-field methods, the interaction energy obtained with the largest basis set 
was directly used, without any extrapolation.
For the explicitly correlated schemes, the Hartree-Fock basis-set incompleteness error 
was reduced by calculating the complete auxiliary basis set singles correction 
(CABS).\cite{adler2007simple,noga2009on}

The large number of individual MBE contributions requires that they are computed 
with high precision.
For Molpro calculations, the orbital and energy convergence criteria 
were set to at least $10^{-8}$ Hartree.
Within MBE, only the F12 calculations used a resolution of identity. 
The AV$n$Z/JKFIT,\cite{weigend2002fully} AV$n$Z/MP2FIT,\cite{weigend2002efficient,weigend2005balanced} 
and AV$n$Z/OPTRI\cite{yousaf2009optimized} 
basis sets were respectively used for exchange fitting, density fitting, and resolution
of identity  (the corresponding Molpro keywords $\tt df\_basis\_exch$, $\tt df\_basis$, and $\tt ri\_basis$). 
Resolution of identity was used also in the HF and MP2 supermolecular calculations.
The cardinal numbers of the auxiliary basis sets were larger by
one compared to the orbital basis sets.

For a further analysis, the contributions of first-order exchange, dispersion, 
and exchange-dispersion energies to the noncovalent interaction energy of dimers
were computed using the SAPT2+3(CCD) scheme\cite{williams1995dispersion,parrish2013tractability} in
the Psi4 program.\cite{parrish2017psi4} The dispersion plus exchange-dispersion energy
was extrapolated using the AVTZ and AVQZ basis sets.
The SAPT exchange term was obtained in the AVQZ basis.
The coupled three-body dispersion energy was computed using
the PBE0 orbitals in the Molpro program and extrapolated
to the CBS limit (AVTZ $\rightarrow$ AVQZ).

The post-Kohn Sham RPA calculations based on PBE, PBE0, SCAN, and SCAN0
were carried out with an in-house code, using a modified version
of the algorithm described in Ref.~\citenum{modrzejewski2020random}
The $n$-body RPA interaction energy can be written down as follows:
\begin{equation}
E_\text{int}^\text{RPA} = E^\text{HF} + E_\text{c}^\text{RPA} \label{rpa-energy-composition}
\end{equation}
where $E^\text{HF}$ is the Hartree-Fock-like energy expression evaluated
on Kohn-Sham orbitals and $E_\text{c}^\text{RPA}$ is the post-Kohn Sham direct
RPA correlation energy. 
All terms on the right-hand side of Eq.~\ref{rpa-energy-composition}
are energy differences computed according to Eqs.~\ref{two-body-energy}--\ref{four-body-energy}.
The RPA interaction energy with the renormalized singles energy\cite{klimes2015singles} (RSE)
is analogous to Eq.~\ref{rpa-energy-composition} but includes the RSE
term
\begin{equation}
E_\text{int}^\text{RPA+RSE} = E^\text{HF} + E_\text{c}^\text{RSE} + E_\text{c}^\text{RPA}
\end{equation}
The RSE term is computed using the density matrix $\Matrix{\rho}^\text{HF}$
obtained from the eigenvectors of the Fock matrix built from
converged Kohn-Sham orbitals:\cite{klimes2015singles}
\begin{multline}
  E_\text{c}^\text{RSE} = \Tr\left( \Matrix{\rho}^\text{HF} \Matrix{F}^\text{HF}\left[\Matrix{\rho}^\text{DFT}\right] \right) \\
  - \Tr\left( \Matrix{\rho}^\text{DFT} \Matrix{F}^\text{HF}\left[\Matrix{\rho}^\text{DFT} \right] \right)
\label{singles-correction}
\end{multline}
The RPA calculations with PBE, PBE0, SCAN, and SCAN0 orbitals employed
the tightest set of numerical precision thresholds
defined in Table~1 of Ref.~\citenum{modrzejewski2020random}.
Unless noted otherwise, the RPA correlation energies were extrapolated to the
complete basis-set limit with the basis sets AVTZ and AVQZ.\cite{halkier1999basis}
As with the other correlated methods, the contribution of the core electrons to the correlation energy was not included.
The Hartree-Fock component and the renormalized singles energy were computed
with the AVQZ basis without any extrapolation.
The method used to optimize the grid for frequency integration
was changed with respect to the previous work.
In Ref.~\citenum{modrzejewski2020random} the quadrature was optimized for a distribution of orbital energy 
differences ($d_{ai}$ in Ref.~\citenum{modrzejewski2020random}) occurring in the full $n$-body 
cluster ({\it e.g.}, a trimer) and all its subsystems.
In this work, the quadrature optimization subroutine ignores the energy differences
which occur in the subsystems but are above the largest occupied-virtual orbital
energy difference in the full $n$-body cluster.
This modification results in a smaller number of quadrature points, especially for tetramers,
while having no discernible effect on numerical precision.
As a result, the number of grid points
is reduced for systems with a large number of ghost atoms.

Special care was taken to avoid numerical errors related to
the numerical integration grid used in the DFT step that
precedes RPA.
It is known, for example, that the SCAN functional is
sensitive to the density of points
employed in numerical integration.\cite{bartok2019regularized,furness2020accurate,mejia2020spin}
Throughout this work we have used a fine grid of
150 radial and 590 spherical points.
The grid was counterpoise-corrected within $n$-body
fragments.
A comparison with the benchmark grid with 250 radial
and 1202 spherical points shows that this choice guarantees
that the grid-related errors are negligible.
In particular, we have checked that the errors in the accumulated
two- and three-body interaction energies at the RPA(SCAN),
RPA(SCAN)+RSE, and DFT(SCAN) levels are on the order
of $10^{-3}$~kcal/mol.
A detailed comparison of the energies computed with the medium, fine,
and extra-fine molecular grids is available in the Supporting Information.

The orbitals and orbital energies generated with the optimized effective potential (OEP)
method, OEPX\cite{talman:1976:OEP} and OEP2-sc,\cite{bartlett2005exchange,UHFOEP2sc}
were computed in the uncontracted correlation-consistent basis sets with
the {\tt ACES II} program.\cite{acesII}
The post-Kohn-Sham RPA calculations based on the OEP orbitals were carried out
with a modified version of the coupled-clusters code of Ref.~\citenum{Smi-gCC-2017}.
Following previous work reported in Refs.~\citenum{OEPSOSszs,smiga2019self,OEPxPBE},
the equations of the OEP method were solved
with the finite-basis set procedure of Refs.~\citenum{gorling:1999:OEP} and~\citenum{ivanov:1999:OEP}.
To calculate the pseudo-inverse of the density-density response matrix, we utilized
truncated singular value decomposition (TSVD).
This step is essential for determining stable and physically meaningful
OEP solutions.\cite{hirata:2001:OEPU,ivanov:2002:OEP,OEPSOSszs}
The cutoff threshold of TSVD was set to $10^{-6}$.
In all OEP calculations, we employed the same tight
thresholds as in Ref.~\citenum{Smi-gCC-2017}.
We refer the reader to Ref.~\citenum{OEPSOSszs} for additional
technical details.

\section{Results}
\subsection{Many-body expansion reference values}

We now discuss the many-body expansion of the interaction energy
at the coupled cluster level.
To obtain reliable reference interaction energy, {\it i.e.}, 
results with the precision of one or two percent, there are several critical issues to overcome.
Most importantly, the values need to be converged with respect to the basis-set size
and also with the order of MBE, {\it i.e.}, the number of molecules in the largest fragment.
Moreover, for large clusters the individual contributions need to be obtained with 
a high precision.\cite{richard2013understanding}

To assess the convergence with the basis-set size, we compare results
obtained with basis-set extrapolation procedures to data computed using explicitly correlated
(F12) methods.
The convergence with the basis-set size is the most critical for dimers 
and, fortunately, the nonadditive contributions of larger clusters converge
much faster with the basis-set size.\cite{gora2011interaction}
Concerning the order of MBE, we computed n-body terms up to clusters containing
four molecules.
To assess the importance of higher-order terms we have compared the results of MBE and
of the supermolecular approach. 
For HF and MP2 they agree within a few hundredths of kcal/mol so that the fourth order
MBE should be sufficiently converged.

\subsubsection{Dimers}
The largest contribution to the interaction energy comes from the twenty
two-body terms involving the interaction between a single water molecule and methane.
The CCSD(T) and CCSD(T)-F12 results are summarized in Table~\ref{tab:cc:2body}.
We present values for all the basis sets used and, where applicable, also estimates
of the basis-set limit obtained with the two-point formula 
of Halkier~{\it et al.}\cite{halkier1999basis}

The total two-body Hartree-Fock contribution shows very little dependence on the basis-set size,
changing by less than 0.01~kcal/mol between AVTZ and AV5Z basis sets.
This is partly due to the cancellation of the basis-set errors of the individual contributions.
The CABS correction has little effect on $E_\text{int}[2]$ but improves the convergence
of the individual terms.
For example, adding CABS reduces the errors in the AVTZ basis set approximately 
by a factor of three.

As expected, the CCSD contribution shows stronger dependence on the basis-set size
so that either extrapolation or the use of the F12 scheme is needed.
The difference between the AVTZ $\rightarrow$ AVQZ and AVQZ $\rightarrow$ AV5Z estimates
of the CCSD correlation energy is 0.028 kcal/mol.
That difference is similar to the difference between CCSD-F12b energies obtained with AVTZ and AVQZ basis sets.
The CCSD-F12b contribution then changes by a mere 0.006 kcal/mol upon going to the AV5Z basis set.
We, therefore, assume that the CCSD-F12b value is closer to the CBS limit.

The perturbative triples (T) term shows also slow convergence with the basis set size 
so that extrapolation to the CBS limit is needed in the canonical approach.
While there are no F12 corrections for triples, the triples term can be scaled with 
the ratio of MP2-F12 and MP2 correlation
energies to achieve faster basis-set convergence.\cite{knizia2009simplified}
Note, however, that there is a subtle point in how this is performed.
One can either scale the monomer (T) energies independently or use the dimer scaling 
factor in all the calculations.
Only the latter approach is size-consistent.\cite{marchetti2009accurate}
In practice, the differences between the approaches are small and for dimers there is no clear
preference for a single scheme.
In particular, there is a close agreement between the two-body (T) energy obtained with independent
scaling and with unscaled (T) extrapolated using the AVQZ and AV5Z basis sets.
This holds also for the individual contributions of the twenty dimers.
Using a common scaling factor leads to a more attractive (T) contribution but there is also a larger
change between the AVQZ basis set and AV5Z basis set compared to the independent scaling approach.
To sum up, independent scaling or extrapolation work well and we take the unscaled (T) obtained by 
extrapolation as the reference value.

The total two-body contribution without and with the use of F12 is $-6.33$ and $-6.31$~kcal/mol,
respectively. 
Based on the basis-set convergence we deem the latter value to be more precise and estimate that 
its uncertainty is 0.01~kcal/mol.
This excludes core correlation and correlation contributions beyond CCSD(T). 
Our value differs from $-5.85$~kcal/mol obtained by Deible and co-workers 
in Ref.~\citenum{deible2014theoretical}.
The reason is the use of the much smaller VTZ-F12 basis set in Ref.~\citenum{deible2014theoretical}.
The HF contributions obtained with the VTZ-F12 and AVTZ basis set differ by less than 0.001 kcal/mol.
However, in the VTZ-F12 basis-set the CCSD-F12b contribution equals $-8.260$~kcal/mol and 
the scaled and unscaled (T) contributions are $-1.544$ and $-1.427$~kcal/mol, respectively.
These values differ considerably from the AVTZ data, see Table~\ref{tab:cc:2body}.
Adding the HF, CCSD-F12b, and unscaled (T) contributions reproduces the value given
in Ref.~\citenum{deible2014theoretical} so that the basis set is indeed the cause of the difference.
The inferior performance of the VTZ-F12 basis set can be most likely attributed to the lack
of diffuse functions.

\begin{table*}[tb]
  \caption{Two-body contributions to the coupled clusters interaction energy (kcal/mol). 
The contributions used in the final coupled-cluster reference are written in bold.}
  \label{tab:cc:2body}
  \begin{threeparttable}
  \begin{tabular}{lrrrrr}
    \toprule
     Method                &  AVTZ  & AVQZ  & AVTZ$\rightarrow$AVQZ & AV5Z & AVQZ$\rightarrow$AV5Z\\
    \midrule
    HF$^a$                 &$3.836$  & $3.832$ &     ---     &  $3.828$& ---\\
    CCSD$^b$               &$-8.175$ & $-8.388$& $-8.543$ & $-8.450$& $-8.515$\\    
    (T)$^c$                &$-1.528$ & $-1.597$& $-1.648$ & $-1.618$& $-1.640$\\
    CCSD(T)$^d$            &$-5.867$ & $-6.153$& $-6.359$ & $-6.240$& $-6.328$ \\
    HF+CABS$^a$            &$3.835$  & $3.830$ &   --- &$\mathbf{3.827}$ & --- \\
    CCSD-F12b$^b$          &$-8.467$ & $-8.494$& --- &$\mathbf{-8.500}$ & ---\\    
    (T)$_{\rm unscaled}^{c}$ &$-1.513$ & $-1.589$& $-1.645$ & $-1.614$ & $\mathbf{-1.639}$\\
    (T)$_{\rm scaled}^{c,e}$   &$-1.624$ & $-1.637$& --- &${-1.638}$ & ---\\
    (T)$_{\rm scaled}^{c,f}$   &$-1.659$ & $-1.655$& --- &${-1.649}$ & ---\\
    CCSD(T)-F12b$^g$       &$-6.144$ & $-6.252$& $-6.308$ & $-6.286$& $-6.312$ \\
    \bottomrule
  \end{tabular}
  \footnotesize{
  \begin{tablenotes}
  \item[$a$] Hartree-Fock contribution to $E_\text{int}[2]$.
  \item[$b$] CCSD correlation-only contribution.
  \item[$c$] Correlation-only contribution of perturbative triples.
  \item[$d$] Sum of HF, CCSD, and (T) values. For AVTZ$\rightarrow$AVQZ and AVQZ$\rightarrow$AV5Z columns, HF data obtained
             with AVQZ and AV5Z basis sets were used, respectively.
  \item[$e$] Independent scaling factors for the dimer and monomers.
  \item[$f$] Common scaling factor for the dimer and monomers.
  \item[$g$] Obtained as a sum of HF+CABS, CCSD-F12b, and (T)$_{\rm unscaled}$ values, 
             for AVTZ$\rightarrow$AVQZ and AVQZ$\rightarrow$AV5Z columns HF+CABS and CCSD-F12b
             values obtained with AVQZ and AV5Z basis sets were used, respectively.
  \end{tablenotes}}
  \end{threeparttable}
\end{table*}

\subsubsection{Trimers}

We now discuss the three-body nonadditive contributions to the interaction energy.
The data for CCSD(T) and CCSD(T)-F12 are summarized in Table~\ref{tab:cc:3body}.
For most of the components, there is almost no dependence on the basis-set size.
For example, the total three-body contribution of canonical CCSD(T) obtained with the AVTZ basis-set 
differs only by around 0.01~kcal/mol from the data extrapolated to the CBS limit.
When the F12 approach is used, CCSD-F12b values in the AVTZ and AVQZ basis sets are within 0.001 kcal/mol.

The only term requiring attention is the triples (T) energy.
The convergence of triples is fast with the basis set
in the unscaled variant and when a common scaling factor is used.
For example, unscaled (T) is essentially converged in the AVTZ basis set. 
By contrast, the convergence is slower when the (T) components
are scaled independently in the monomer and dimer subsystems.

Overall, the reference total three-body contribution using the CCSD(T) method is 1.04~kcal/mol.
The values taken to obtain this value are in bold in Table~\ref{tab:cc:3body}.
Our result agrees with the value of $E_\text{int}[3]=1.01$~kcal/mol
reported  by Deible {\it et al.},\cite{deible2014theoretical} which again demonstrates 
the weaker dependence of the three-body term on the basis-set size.

\begin{table*}[tb]
  \caption{Three-body contributions to the coupled clusters interaction energy (kcal/mol).
The contributions used in the final coupled-cluster reference are written in bold.}
  \label{tab:cc:3body}
  \begin{threeparttable}
  \begin{tabular}{lrrr}
    \toprule
     Method  &  AVTZ  & AVQZ  & AVTZ$\rightarrow$AVQZ \\
    \midrule
      HF$^a$                   &$-0.273$  & $-0.273$ &  ---   \\
      CCSD$^b$                 &$1.112$ & $1.108$& $1.106$ \\
      (T)$^c$                  &$0.205$ & $0.203$& $0.202$ \\
      CCSD(T)$^d$              &$1.044$ & $1.039$& $1.035$ \\ 
      HF+CABS$^a$              &$-0.273$  & $\mathbf{-0.272}$ & ---  \\
      CCSD-F12b$^b$            &$1.106$ & $\mathbf{1.106}$&  ---\\
      (T)$_{\rm unscaled}^c$ &$0.201$ & $0.202$& $\mathbf{0.202}$ \\
      (T)$_{\rm scaled}^{c,e}$   &$0.046$ & $0.114$& --- \\
      (T)$_{\rm scaled}^{c,f}$   &$0.221$ & $0.210$& --- \\
      CCSD(T)-F12b$^g$         &$1.034$ & $1.035$& $1.035$ \\
    \bottomrule
  \end{tabular}
  \footnotesize{
  \begin{tablenotes}
  \item[$a$] Hartree-Fock contribution to $E_\text{int}[3]$.
  \item[$b$] CCSD correlation-only contribution.
  \item[$c$] Correlation-only contribution of perturbative triples.
  \item[$d$] Sum of HF, CCSD, and (T) contributions, value in AVTZ$\rightarrow$AVQZ column 
             uses HF data obtained with AVQZ basis set.
  \item[$e$] (T) contributions scaled individually in each calculation.
  \item[$f$] Common scaling factor for (T) contributions.
  \item[$g$] Sum of HF+CABS, CCSD-F12b, and (T)$_{\rm unscaled}$ data, in AVTZ$\rightarrow$AVQZ column 
             the HF+CABS and CCSD-F12b results obtained with AVQZ basis set are used.
  \end{tablenotes}}
  \end{threeparttable}
\end{table*}

\subsubsection{Tetramers}

There are 1140 tetramers, many more compared to the 20 dimers and 190 trimers, suggesting
a much larger computational effort.
However, given that the total three-body term differs by only 0.001~kcal/mol between the AVTZ
and AVQZ basis sets, we expect that the four-body term will show similar behavior. 
We have therefore used the CCSD(T)-F12b scheme together with the AVTZ basis-set to obtain
the four-body terms.
Spot checks on twenty tetramers show that for CCSD-F12b the largest differences between 
the AVTZ and AVQZ data are on the order of 10$^{-5}$~kcal/mol for a single tetramer.
Since there are 1140 tetramers, the difference upon going to the AVQZ basis could be at most
on the order of $10^{-2}$~kcal/mol. 
However, we expect it to be smaller as some of the individual contributions are positive 
and some negative.
The weak dependence on the basis set size is also confirmed by tetramer calculations 
in the AVDZ basis set.

The CCSD(T)-F12b four-body energy and the contributions of the different components 
are shown in Table~\ref{tab:cc:4body}.
The Hartree-Fock term dominates over the correlation part.
As with the three-body terms, individual scaling of the triples should be avoided.
Using a common scaling factor or doing no scaling of triples leads to a fast convergence with the 
basis set size.

\begin{table}[tb]
  \caption{Four-body contributions to the coupled clusters interaction energy, data in kcal/mol.
The contributions used in the final coupled-cluster reference are written in bold.}
  \label{tab:cc:4body}
  \begin{threeparttable}
  \begin{tabular}{lrr}
    \toprule
     Method  &  AVDZ & AVTZ \\
    \midrule
      HF+CABS$^a$    &${ 0.548}$          & ${\bf 0.552}$  \\
      CCSD-F12b$^b$  &${ -0.010}$         & ${\bf -0.016}$ \\
      (T)$_{\rm unscaled}^c$&${ 0.019}$   & ${\bf 0.021}$\\
      (T)$_{\rm scaled}^{c,d}$  &$-1.480$ & $0.895$\\
      (T)$_{\rm scaled}^{c,e}$  &$0.025$  & $0.023$\\
      CCSD(T)-F12b$^{f}$ &${ 0.557}$      & $ 0.557$  \\
    \bottomrule
  \end{tabular}
  \footnotesize{
  \begin{tablenotes}
  \item[$a$] Hartree-Fock contribution to $E_\text{int}[4]$.
  \item[$b$] CCSD correlation-only contribution.
  \item[$c$] Correlation-only contribution of perturbative triples.
  \item[$d$] (T) contributions scaled individually in each calculation.
  \item[$e$] Common scaling factor for (T) contributions.
  \item[$f$] Sum of HF+CABS, CCSD-F12b, and (T)$_{\rm unscaled}$ contributions.
  \end{tablenotes}}
  \end{threeparttable}
\end{table}

The CCSD(T) interaction energy obtained up-to fourth order of MBE is $-4.72$~kcal/mol.
We estimate the higher-order effects as the difference between supermolecular and MBE calculation 
at the Hartree-Fock level, which is 0.01~kcal/mol.
Therefore, our final estimate of the CCSD(T) interaction energy between methane 
and the water cage is $-4.71$~kcal/mol. 
We estimate its numerical uncertainty to be below one percent or 0.04~kcal/mol.
The value nor its uncertainty include core correlations or correlations beyond the (T) term.
The uncertainty comes from basis-set convergence, possible loss of precision of the four-body 
terms,\cite{richard2013understanding} and five-body and higher correlation contributions.

\subsubsection{Comparison to previous benchmark results}

We now discuss the differences between the hereby presented reference value and
other estimates published in the literature.
Lao and Herbert\cite{lao2018simple} employed the domain-based local pair natural orbital CCSD(T)
approach (DLPNO-CCSD(T))\cite{riplinger2013natural,liakos2015exploring}
and obtained $E_\text{int}=-4.88$~kcal/mol in a supermolecular calculation.
The difference between that value and our result originates from
the basis-set extrapolation method employed in Ref.~\citenum{lao2018simple}
(MP2/CBS + $\delta_\text{CCSD(T)}/\text{def2-TZVPP}$) and the approximations
inherent in the DLPNO scheme.\cite{liakos2015exploring}
Deible~{\it et al.}\cite{deible2014theoretical} obtained
$E_{\rm int}=-5.3\pm0.5$~kcal/mol using diffusion Monte Carlo (DMC). 
The uncertainty of $0.5$~kcal/mol includes the stochastic uncertainty 
of the DMC energy and uncertainty due to extrapolation to zero time step.
%
Additional uncertainty comes from the fixed-node approximation and the choice
of the Jastrow factor.\cite{dubecky2019toward}
A part of the difference between our value and that of Deible~{\it et al.}\cite{deible2014theoretical}
originates from the electron correlations beyond the CCSD(T) level, as these are captured by DMC.
For a water-methane dimer, \v{R}ez\'{a}\v{c} and co-workers found interaction energy difference 
between CCSDT(Q) and CCSD(T) to be $-0.006$~kcal/mol.\cite{rezac2015extensions}
This would translate to a difference of $-0.12$~kcal/mol for the twenty dimers in the two-body term
and a reduction of the difference between our value and the DMC data.
Finally, none of the results mentioned include the effect of core correlation.
\v{R}ez\'{a}\v{c}~{\it et al.} found core correlations to contribute $-0.006$~kcal/mol
for a water-methane dimer,\cite{rezac2015extensions} which would possibly mean a total effect of $-0.12$~kcal/mol
when summed over twenty dimers.
However, our MP3 calculations employing the aug-cc-pwCVQZ basis set\cite{dunning1989gaussian,kendall1992electron,peterson2002accurate} 
show that the sum of two-body and nonadditive three-body core-electron contributions is
only $-0.02$~kcal/mol ($E_\text{int}^\text{core}[2]=-0.028$ and $E_\text{int}^\text{core}[3]=0.006$~kcal/mol).
The core-correlation term is not included in the reference $E_\text{int}$
further in the text.

\subsubsection{M{\o}ller-Plesset perturbation theory}

Before we use the coupled cluster benchmark to assess RPA, 
we apply the M{\o}ller-Plesset hierarchy of methods to gain additional insight
into the energy contributions.
Both MP2 and MP3 underestimate the two-body term 
by approximately 1~kcal/mol (Table~\ref{tab:cc:NN}).
MP4 performs much better and is within 0.05~kcal/mol of the reference value.
There are two types of trimers, with and without hydrogen bonded water molecules,
we denote them ``3b 1hb'' and ``3b 0hb'', respectively.
As discussed by Deible~{\it et al.}, MP2 lacks three-body dispersion 
contributions and the total three-body term is too attractive for either of the groups.
The MP3 and MP4 results are close to the reference values both for the 
hydrogen-bonded and non-hydrogen-bonded trimers.
We note that a simple arithmetic mean of the MP2 and MP3 interaction energies (the MP2.5 approach) has been 
proposed to improve the interaction energies of dimers and nonadditive three-body energies of non-covalent 
clusters.\cite{pitonak2009scaled,rezac2015benchmark}
Interestingly, this scheme would not improve the accuracy of MP3 for the three-body terms
in clathrate.

The largest contributions to four-body correlation energy come from the clusters with 
two hydrogen bonds.
Here MP2 does not provide enough attractive correlation leading to too repulsive total 
four-body contribution (by 0.06~kcal/mol).
MP3 does not bring an overall improvement compared to MP2, the errors are reduced for the 4b~2hb subset
but the accuracy deteriorates for the 4b~1hb fragments.
Interestingly, the total HF+CABS four-body contribution is within 0.01~kcal/mol of the CCSD(T) reference,
which is a result of a fortunate cancellation of errors.

The total sum of the two-, three-, and four-body terms shows that both MP2 and MP3 underestimate
the total interaction energy, mostly due too weakly binding two-body contributions.
The aforementioned lack of three-body dispersion in MP2 partly cancels the error of two-body terms so that 
the overall error of MP2 ($0.39$~kcal/mol) is about one half of the MP3 one ($0.91$~kcal/mol).
Approximating the four-body MP4 terms by HF values, we estimate its interaction energy to be approximately 
$-4.83$~kcal/mol, around 0.1~kcal/mol away from the reference value.

Interestingly, the trend of the MP2 errors for the three-body contributions is consistent with the 
Axilrod-Teller-Muto (ATM) formula for three-body dispersion.\cite{Axilrod-Teller:3Bdispersion,muto1943force}
According to the ATM term, the three-body dispersion is attractive for linear configurations
and repulsive for trimer angles below approximately 117\si{\degree}.
The 3b 1hb configurations have a structure similar to an isosceles triangle with a vertex angle around 40\si{\degree} 
and the ATM three-body dispersion term is repulsive.
For these trimers, MP2 recovers only around 50~\% of the (repulsive) CCSD(T) correlation.
For the same reason MP2 also overbinds the proximate 0hb trimers (formed by second nearest neighbor water molecules).
With increasing distance between the water molecules the trimer angle increases so that
the error is around zero for third nearest neighbor waters and positive for trimers close to linear geometry, 
see SI and additional resources.\cite{klimes2020git}

\begin{table*}
  \setlength{\tabcolsep}{5pt}
        \caption{Contributions to the methane--water cage interaction energy divided into $n$-body terms ($n$b)
        and subsystems with $m$ hydrogen bonds ($m$hb). Energies are in kcal/mol.}
        \label{tab:cc:NN}
\begin{threeparttable}
\begin{tabular}{lrrrrrrrrr}
\toprule
Method  &  2b & 3b 1hb & 3b 0hb & 4b 2hb & 4b 1hb & 4b 0hb & Sum\\
\midrule
HF+CABS$^a$ &     $3.827$ & $-1.483$ & $1.210$ & $0.164$ & $0.471$ & $-0.083$ & $4.107$\\
CCSD-F12b$^b$  &  $-8.500$ & $0.913$ & $0.193$ & $-0.093$ & $0.054$ & $0.023$& $-7.410$ \\
(T)$^c$     &    $-1.639$ & $0.172$ & $0.030$ & $-0.015$ & $0.027$ & $0.009$&  $-1.415$ \\
CCSD(T)-F12b$^d$ &    $-6.312$ & $-0.398$ & $1.434$ & $0.056$ & $0.552$ & $-0.051$& $-4.718$ \\
\midrule
MP2$^e$     &  $-5.192$ & $-0.956$ & $1.208$ & $0.117$ & $0.563$ & $-0.067 $&$-4.326$\\
MP3$^e$     &  $-5.452$ & $-0.382$ & $1.501$ & $0.077$ & $0.519$ & $-0.067 $&$-3.803$\\
MP4$^e$     &  $-6.360$ & $-0.433$ & $1.409$ & ---     & ---     & ---      & --- \\
\bottomrule
\end{tabular}
\footnotesize{\begin{tablenotes}
\item[$a$] Hartree-Fock contribution to $E_\text{int}[n\text{b}\;m\text{hb}]$.
\item[$b$] CCSD-F12b correlation-only contribution.
\item[$c$] Correlation-only contribution of perturbative triples, no scaling of triples used.
\item[$d$] Sum of HF+CABS, CCSD-F12b, and (T) energy terms.
\item[$e$] $E_\text{int}^{\text{MP}n} = E^\text{HF} + E_\text{c}^{\text{MP}n}$.
\end{tablenotes}}
\end{threeparttable}
\end{table*}

\subsubsection{Basis sets for fragment calculations}
\label{sec:bsse}

There is an ample body of literature devoted to the choice of basis sets for
MBE calculations.\cite{valiron1997hierarchy,walczak2011on,richard2013achieving,ouyang2015many,liu2017understanding,richard2018understanding,peyton2019basis,gora2011interaction}
A particular question of interest is whether the fragment-centered basis set
is sufficient for $n$-body contributions or if basis functions on other atoms 
need to be considered as well.
As already noted by Góra~and co-workers,\cite{gora2011interaction} at the CBS limit, 
the full cluster and fragment basis sets yield identical results.
In the following we compare a single dimer and a single trimer contribution to show that
this is the case also for the methane clathrate and there is no error caused by the use of the fragment basis set.

\begin{figure}
  \includegraphics[width=0.45\textwidth]{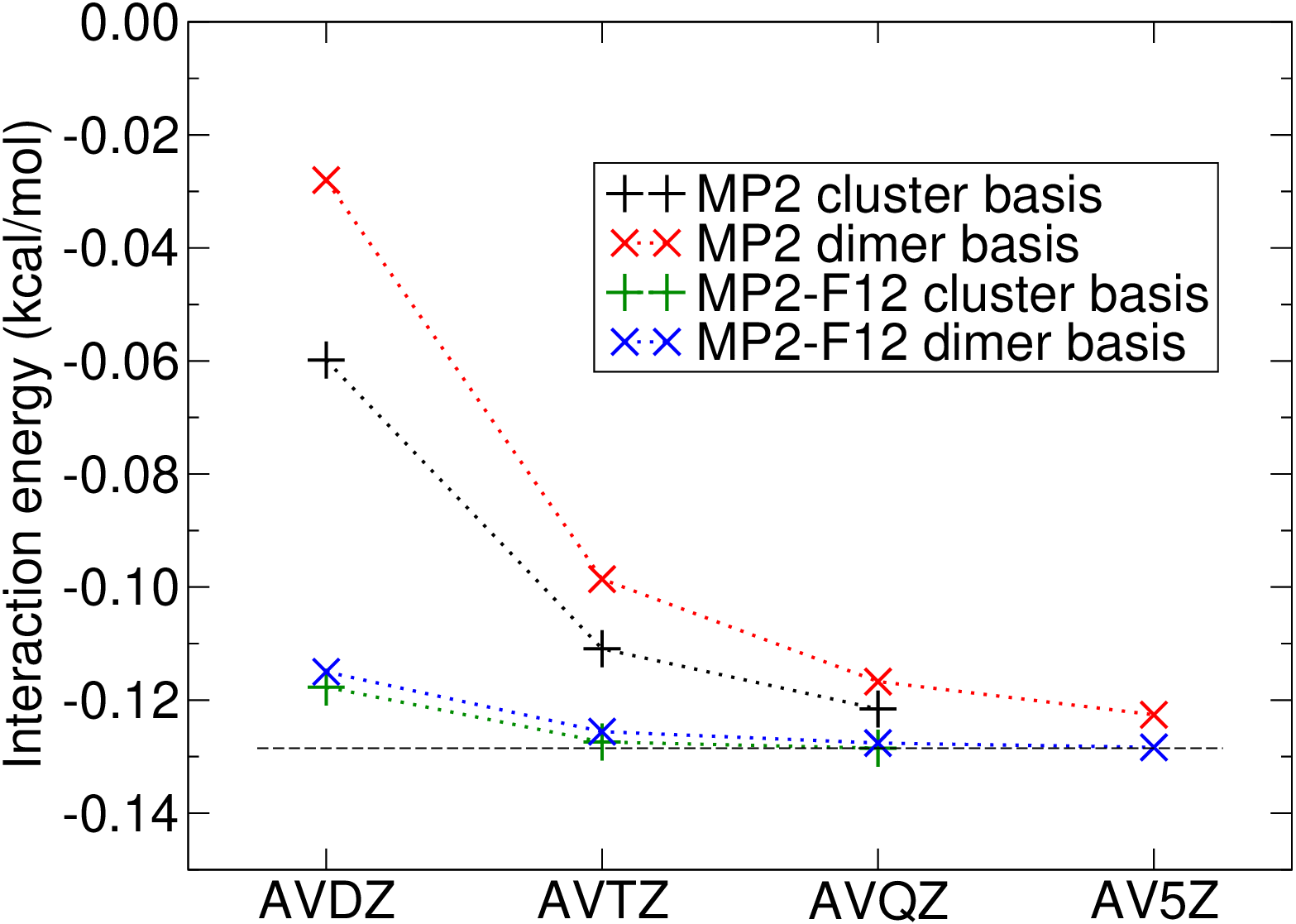}
  \caption{Interaction energy of water-methane dimer (water 01) obtained
using basis set functions at every atom of \ce{CH4(H2O)20} (cluster basis) and basis set with functions
centered on the interacting dimer only (dimer basis).}
  \label{bsse:dimer}
\end{figure}

\begin{figure}
  \includegraphics[width=0.45\textwidth]{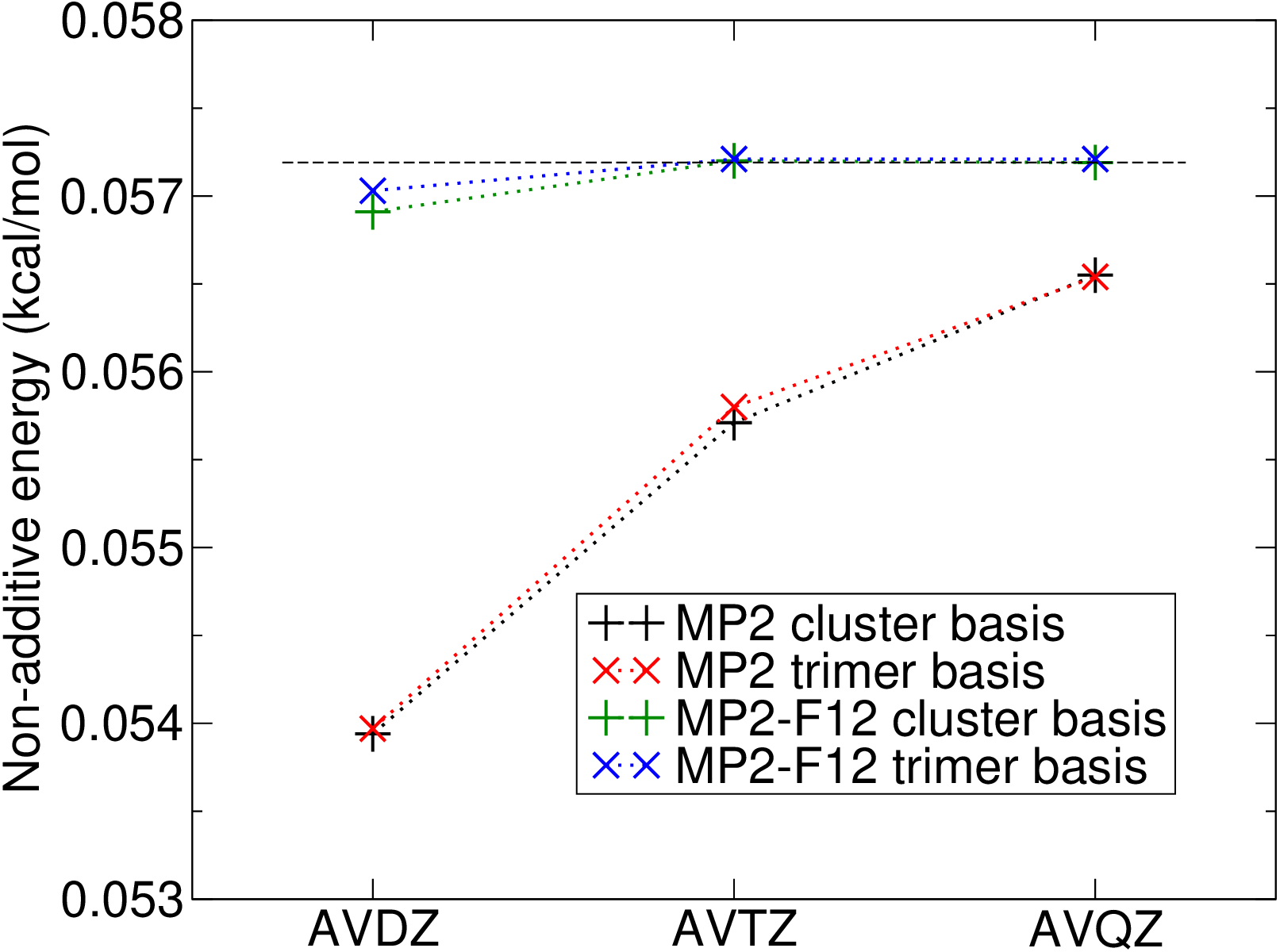}
  \caption{Three-body nonadditive energy of a 3b 1hb trimer (water 01, water 02, methane)
obtained using the full cluster basis and the trimer basis.}
  \label{bsse:trimer}
\end{figure}

MP2 and MP2-F12 dimer interaction energies in the dimer and full-cluster basis sets,
with increasing cardinal numbers, are shown in Fig.~\ref{bsse:dimer}.
Clearly both basis set types lead to the same CBS limit.
The rate of convergence of the full-cluster energies is only marginally faster, 
while the calculations are much more computationally demanding.
The difference between the full-cluster basis and fragment basis becomes
even less important for the three-body contribution, as demonstrated in Fig.~\ref{bsse:trimer}.
Here the selected trimer belongs to the 3b~1hb subset.
The difference between the MP2-F12 nonadditive energies in the full-cluster
and trimer basis set is $< 2\cdot10^{-4}$~kcal/mol in the AVDZ basis set
and an order of magnitude smaller than that for the AVTZ and AVQZ basis sets.
We conclude that the $n$-body fragment-centered basis sets
employed in this work for all explicit MBE terms do not introduce any 
considerable basis set error and require less computational effort compared to 
the cluster basis set.

\subsection{Performance of RPA}
\subsubsection{Supermolecular approach}
For clusters and molecular solids,
the RPA energy can be obtained in a single step
without resorting to the many-body expansion, using one of efficient
RPA algorithms.\cite{eshuis2010fast,delben2013electron,kaltak2014low,kaltak2014cubic,schurkus2016effective,wilhelm2016large,klimes2016lattice,modrzejewski2020random}
The total supermolecular methane---water cage interaction energies are in
Table~\ref{total-rpa-energies}.
The energies are obtained using RPA based on pure DFT functionals
(PBE and SCAN) and their hybrid variants with $25~\%$ exact exchange
(PBE0 and SCAN0).
As expected, RPA without RSE is underbinding by 1 to 1.5~kcal/mol regardless
of the exchange-correlation model.
With RSE included, the results split into two groups separated
by ca.~1~kcal/mol:
overbinding RPA+RSE based on pure functionals and underbinding RPA+RSE based
on hybrids.
For the total supermolecular interaction energy, RPA(PBE)+RSE
and RPA(SCAN)+RSE appear to be the best performing methods.
However, as will be evident from the many-body expansion,
the errors in the supermolecular energy are dominated by
the dimer contribution and hide the erratic behavior of
some of the methods in the nonadditive $n$-body terms.

\begin{table}[tb]
  \setlength{\tabcolsep}{5pt}
  \caption{Total methane---water cage interaction energy (kcal/mol).$^a$}
  \label{total-rpa-energies}
  \begin{threeparttable}
  \begin{tabular}{lrrrr}
    \toprule
             & SCAN0   &  PBE0  &  SCAN & PBE \\
    \midrule
    RPA+RSE & $-4.18$ & $-4.03$ & $-5.11$ & $-4.99$ \\
    RPA   & $-3.17$ & $-3.30$ & $-3.38$ & $-3.73$ \\
    DFT$^e$   & $-3.08$ & $1.01$ & $-4.04$ & $1.29$ \\
    \midrule
    HF+CABS$^b$      & \multicolumn{4}{c}{$4.12$} \\
    MP2-F12$^b$     & \multicolumn{4}{c}{$-4.30$} \\
    CCSD(T)$^c$ & \multicolumn{4}{c}{$-4.71$} \\
    DMC$^d$              & \multicolumn{4}{c}{$-5.3\pm0.5$} \\
    \bottomrule
  \end{tabular}
  \footnotesize{
  \begin{tablenotes}
  \item[$a$] RPA correlation energies are extrapolated to the basis set limit (AVTZ$\rightarrow$AVQZ).
    $E^\text{EXX}$, $E_\text{c}^\text{RSE}$, and self-consistent field DFT energies are
    computed with the AVQZ basis.
  \item[$b$] Supermolecular calculations of the HF and MP2-F12 energies employ the AVQZ orbital basis
  and augmented five-zeta auxiliary basis sets. 
  \item[$c$] Includes the Hartree-Fock estimate of $n>4$-body terms (0.01~kcal/mol).
  \item[$d$] Diffusion Monte Carlo, Ref.~\citenum{deible2014theoretical}.
  \end{tablenotes}
  }
  \end{threeparttable}
\end{table}

\subsubsection{Dimers}

The dimer interactions of \ce{CH4\bond{...}H2O} are challenging
because the interaction energy has the characteristics
of dispersion dominated systems.
In the SAPT2+3(CCD) decomposition,\cite{parrish2013tractability}
almost the entire binding comes from the dispersion contribution
($E_\text{disp}+E_\text{exch-disp}=-11.55$~kcal/mol), which 
is largely balanced by the first-order exchange repulsion energy
($E_\text{exch}=8.91$~kcal/mol).
As we have found in our previous work, in such systems the RPA interaction
energy tends to be sensitive to the choice of
the Kohn-Sham determinant.\cite{modrzejewski2020random}

Let us analyze the accumulated 20 interaction energies of dimers
in Table~\ref{dimer-rpa-energies}.
(The individual dimer energies are available in the Supporting Information.)
The repulsive interaction is accounted for in the Hartree-Fock-like
energy contribution $E^\text{HF}$.
Among pure and hybrid (meta-)GGAs, $E^\text{HF}$ varies by over 2~kcal/mol,
which is significant on the scale of the total interaction energy.
In all cases, the mean field contribution evaluated
on DFT orbitals is more repulsive than
the self-consistent Hartree-Fock energy,
by 1.5~kcal/mol to almost 3.9~kcal/mol.
The agreement between different functionals improves once
the singles correction, $E_\text{c}^\text{RSE}$, is added to
the mean field contribution:
the sum $E^\text{HF}+E_\text{c}^\text{RSE}$
falls within a narrow range of 4.3--4.9~kcal/mol.
Importantly, this is still 0.5 to 1.0~kcal/mol more repulsive
than the interaction energy at the self-consistent
Hartree-Fock level ($E^\text{HF}=3.83$~kcal/mol).

The RPA correlation contribution for pure functionals is visibly larger than for hybrids,
which is due to a smaller electronic gap obtained with the former methods.
For RPA(SCAN)+RSE and RPA(PBE)+RSE, this leads to an overbinding in
the total two-body energy, by 0.15 and 0.39~kcal/mol, respectively.
By contrast, the methods based on hybrids with 25\% of exact exchange,
RPA(SCAN0)+RSE and RPA(PBE0)+RSE, underestimate the two-body energy
by as much as 1~kcal/mol.

Obviously the two sources of errors in direct RPA
are approximate Kohn-Sham eigenstates and
the lack of the exchange-correlation kernel.
It is worthwhile to investigate what level of accuracy can
be achieved if one stays within the direct RPA framework,
but uses high-quality orbitals and orbital energies.
While the main focus remains on the PBE/SCAN-based exchange-correlation
potentials, we briefly introduce a sequence of RPA variants based on ab
initio DFT\cite{bartlett2010ab} orbital inputs:
from Hartree-Fock, to the exchange-only optimized effective potential (OEPX),
to the optimized effective potential with second order
correlation (OEP2-sc).

From the difference of $E_\text{int}^\text{RPA(OEP2-sc)+RSE}$ and $E_\text{int}^\text{RPA(OEPX)+RSE}$,
we estimate that the appropriate treatment of electron correlation in the input 
orbitals contributes  as much as 2~kcal/mol or 30~\% of the total two-body energy
in the methane clathrate cluster (Table~\ref{dimer-rpa-energies}).

Most importantly, we find that the RSE correction
in RPA based on ab initio DFT is still large and necessary to
correct for underbinding.
For example, while $E_\text{c}^\text{RSE}$ is zero in RPA(HF) by definition,\cite{klimes2015singles}
the singles correction is as large as 1.7~kcal/mol for OEPX and
2.4~kcal/mol for OEP2-sc orbitals.
If we take RPA(OEP2-sc) as the reference, the mean-field contribution,
$E^\text{HF}+E_\text{c}^\text{RSE}$, is significantly better reproduced by
(hybrid) GGAs and meta-GGAs than by the Hartree-Fock orbitals.
As seen in Table~\ref{dimer-rpa-energies}, the RPA correlation
component of the interaction energy, $E_\text{c}^\text{RPA}$, becomes more attractive as one progresses
from the Hartree-Fock reference to OEPX to OEP2-sc,
in parallel with the narrowing HOMO-LUMO orbital energy gap.

\begin{table*}[bt]
  \caption{Total interaction energies and interaction energy components of 20 dimers \ce{CH4\bond{...}H2O}.
  Highest occupied orbital energies and LUMO-HOMO energy differences of various DFT methods,
  averaged over 20 dimers. The units of interaction energies and orbital energies are kcal/mol and eV, respectively.}
  \label{dimer-rpa-energies}
  \begin{threeparttable}
  \begin{tabular}{lrrrrrrr}
    \toprule
             & SCAN0$^a$   &  PBE0$^a$  &  SCAN$^a$ & PBE$^a$ & HF$^a$ & OEPX$^b$ & OEP2-sc$^b$ \\
    \midrule
    DFT         & $-4.22$ &  $-2.37$        & $-5.17$     & $-3.88$   & $3.83$  &          &             \\
    RPA+RSE$^c$ & $-5.36$ & $-5.41$ & $-6.46$ & $-6.70$ & $-2.98$ & $-4.47$ & $-6.37$ \\
    RPA$^d$ & $-4.35$ & $-4.14$ & $-4.43$ & $-3.87$ & $-2.98$ & $-2.77$ & $-3.94$ \\
    CCSD(T) & \multicolumn{7}{c}{$-6.31$} \\
    \multicolumn{8}{c}{} \\
    $E_\text{c}^\text{RPA}$ & $-9.74$ & $-9.77$ & $-11.17$ & $-11.56$ & $-6.81$ & $-8.65$ & $-11.04$ \\   
    $E^\text{HF}+E_\text{c}^\text{RSE}$ & 4.38 & 4.36 & 4.71 & 4.87 & 3.83 & 4.18 & 4.66 \\
    $E^\text{HF}$ & 5.39 & 5.62 & 6.74 & 7.69 & 3.83 & 5.88 & 7.10 \\
    \multicolumn{8}{c}{} \\
    & \multicolumn{7}{c}{\ce{CH4}} \\
    \multicolumn{8}{c}{} \\
    $\epsilon_\text{HOMO}$$^e$ & $-11.3$ & $-11.0$ & $-9.9$ & $-9.5$ & $-14.9$ & $-14.9$ & $-14.0$ \\
    ${\epsilon_\text{LUMO}-\epsilon_\text{HOMO}}^f$ & 11.7 & 11.1 & 10.2 & 9.1 & 15.6 & 10.3 & 10.0 \\
    \multicolumn{8}{c}{} \\
    & \multicolumn{7}{c}{\ce{H2O}} \\
    \multicolumn{8}{c}{} \\    
    ${\epsilon_\text{HOMO}}^e$ & $-9.3$ & $-9.0$ & $-7.6$ & $-7.2$ & $-13.8$ & $-13.8$ & $-11.2$ \\
    ${\epsilon_\text{LUMO}-\epsilon_\text{HOMO}}^f$ & 9.3 & 8.6 & 7.2 & 6.2 & 14.4 & 8.2 & 6.8 \\
    \bottomrule
  \end{tabular}
  \footnotesize{
  \begin{tablenotes}
  \item[$a$] CBS extrapolation (AVTZ$\rightarrow$AVQZ) for the correlation part
  of the interaction energy; the AVQZ basis for orbital energies.
  \item[$b$] CBS extrapolation (aug-cc-pVDZU$\rightarrow$aug-cc-pVTZU) for interaction energies
  and their components; the aug-cc-pVTZU basis for orbital energies. ``U'' denotes an uncontracted
  basis set.
  \item[$c$] Label RPA+RSE denotes the interaction energy $E_\text{int}^\text{RPA+RSE} = E^\text{HF} + E_\text{c}^\text{RSE} + E_\text{c}^\text{RPA}$
  \item[$d$] Label RPA denotes the interaction energy $E_\text{int}^\text{RPA} = E^\text{HF} + E_\text{c}^\text{RPA}$
  \item[$e$] $-\epsilon_\text{HOMO}$ approximates the lowest vertical ionization potential;\cite{yang2012derivative}
    experimental IPs for water and methane molecules are 12.6~eV and 13.6~eV, respectively.\cite{linstrom2013nist} 
  \item[$f$] Note that the HOMO-LUMO gap is a well-defined approximation of the lowest excitation energy
    in GGAs and OEP methods; this no longer holds for functionals that include the Hartree-Fock exchange
    operator.\cite{van2014physical,smiga2019self,smiga2016self}
  \end{tablenotes}
  }
  \end{threeparttable}
\end{table*}

\subsubsection{Trimers}

The three-body nonadditive energies tend to be described inaccurately by 
DFT approximations and the clathrate cluster is not an exception.
Indeed, Table~\ref{rpa-trimers} shows that at the self-consistent DFT level the energies
behave erratically, for SCAN and SCAN0 the three-body energies are attractive,
for PBE and PBE0 they are strongly repulsive.
Similar observations have been made by Deible~{\it et al.} in their study.\cite{deible2014theoretical}
Therefore, it is interesting to inspect to what extent is the post-Kohn-Sham RPA 
sensitive to the inaccuracies in the approximate DFT Hamiltonians.  

The RPA's sensitivity to input eigenstates is reflected in the behavior of
the $E^\text{HF}$ component of the RPA interaction energy.
Unlike the self-consistent DFT energy, $E^\text{HF}$ is negative
for all orbital inputs, but still varies in magnitude between
different models.
$E^\text{HF}$ is negative but close to zero for SCAN0 and equals $-4.3$~kcal/mol for PBE.
While it is the total RPA interaction energy that has a direct
physical interpretation, the variance of $E^\text{HF}$ due to
the choice of the semilocal exchange model is an indicator
of the quality of the exchange potential model.
The variance can be assessed, for example, by comparing the results obtained with pure
functional and its hybrid variant.
In this regard, the effect is particularly large for the PBE functional, 
the $E^\text{HF}$ components of RPA(PBE) and RPA(PBE0) differ by 3~kcal/mol.
By contrast, the $E^\text{HF}$ components of RPA(SCAN)
and RPA(SCAN0) are close to each other, which indicates that SCAN orbitals
are more robust against the addition of exact exchange.

Adding now the $E^\text{HF}$ and $E_\text{c}^\text{RPA}$ components together we find that 
the RPA(PBE) and RPA(PBE0) three-body nonadditive interaction energies have incorrect signs.
Moreover, the nonadditive energy of RPA(PBE) exhibits an error of almost 3~kcal/mol.
Those errors are most likely artifacts of the PBE model.
The total three-body nonadditive interaction energies of RPA(HF) are on par with 
RPA(SCAN) and RPA(SCAN0), underestimating the reference value by around 0.5~kcal/mol.

The RSE-corrected approaches RPA(PBE)+RSE and RPA(PBE0)+RSE
are free from the issues described above.
For the SCAN-based variants, the singles corrections are small and slightly
deteriorate the results.
That is consistent with our observations in Ref.~\citenum{modrzejewski2020random},
where RSE was found to slightly worsen the RPA(SCAN0) energies on the dataset
of Řezáč et al.\cite{rezac2015benchmark}
The results shown here further support the previous recommendation not to use 
RSE with RPA(SCAN0) when computing three-body energies separately.

Finally, note that the best agreement for the total three-body term is given by 
RPA(PBE)+RSE.
This is true both for trimers with hydrogen-bonded water molecules (subset 3b 1hb) 
and trimers with no hydrogen bond (subset 3b 0hb), see Table~\ref{rpa-all-subsets}.
However, this is rather deceptive, especially for the 3b 0hb subset, where there
are both positive and negative errors that mostly cancel out.
The mean absolute error (MAE) of RPA(PBE)+RSE for the 3b 0hb subset (equal to 0.0023~kcal/mol)
is about twice as large as the MAE of 0.001~kcal/mol obtained both for RPA(PBE0)+RSE and RPA(SCAN0).
Also RPA(SCAN) and RPA(HF) show lower MAE values for the 3b 0hb subset, 0.0015 and 0.0017~kcal/mol,
respectively.

\begin{table*}[htb]
  \setlength{\tabcolsep}{5pt}
  \caption{Nonadditive interaction energies (kcal/mol) of 190 trimers \ce{CH4(H2O)2}.}
  \label{rpa-trimers}
  \begin{threeparttable}
  \begin{tabular}{lrrrrr}
  \toprule
                                   & SCAN0 & PBE0 & SCAN & PBE    & HF \\
\midrule
  RPA+RSE                          & 0.40 & 0.74 & 0.41 & 0.94   & 0.53 \\
  RPA                              & 0.64 & $-0.21$ & 0.59 & $-2.01$     & 0.53 \\
  DFT/HF$^a$                       & $-0.11$ & 4.39 & $-0.54$ & 7.45 & $-0.27$\\
  CCSD(T) & \multicolumn{5}{c}{1.04} \\
  \multicolumn{6}{c}{} \\
    $E_\text{c}^\text{RPA}$            & 0.66  & 1.46 & 0.64 & 2.34 & 0.81 \\   
  $E^\text{HF}+E_\text{c}^\text{RSE}$ & $-0.26$ & $-0.71$ & $-0.23$ & $-1.40$ & $-0.27$ \\
  $E^\text{HF}$                     & $-0.01$ & $-1.67$ & $-0.04$ & $-4.35$ & $-0.27$ \\
  \bottomrule
\end{tabular}
\footnotesize{
\begin{tablenotes}
\item[$a$] Self-consistent field energies are obtained with the AVQZ basis.
\end{tablenotes}}
\end{threeparttable}
\end{table*}

\subsubsection{Tetramers and $n>4$-body clusters}

The MBE of the methane clathrate is typical of a nonpolar system.
Its terms decay fast with the number of interacting molecules $n$.
The Hartree-Fock contribution dominates the four-body term and amounts to $0.55$~kcal/mol.
The five- and higher-body contributions sum up to almost zero at the Hartree-Fock level.
The CCSD(T) correlation energy contributes less than 0.01~kcal/mol to 
the four-body nonadditive energy; we assume it to be negligible for $n>4$.

Taking into account the limited role of electron correlation, it is expected that
RPA performs well for $n \ge 4$.
Indeed, the best variants overall, RPA(PBE0)+RSE and RPA(SCAN0),
deviate from the reference by less than $0.1$~kcal/mol for the four-body energy 
and predict vanishingly small $n>4$ contributions (Table~\ref{four-body-rpa-energies}).
Therefore, they are consistent with the decay rate of $n$-body terms shown by the CCSD(T) scheme.

It may appear puzzling why some of the other RPA variants, {\it e.g.}, RPA(PBE) and RPA(PBE0),
overshoot the four-body energy by more than a factor of two and show large residuals
in the $n>4$ terms.
For those methods, the MBE is still not converged after including the four-body terms.
A compelling explanation is that the error stems from the inaccurate description of water at the PBE level
and becomes partly inherited by RPA.
In the self-consistent PBE description, the four-body energies have the wrong sign and a magnitude which 
is several times too large (Table~\ref{four-body-rpa-energies}).
As reported by Chen {\it et al.},\cite{chen2017ab} PBE overestimates the polarizability of water,
which results in bulk waters having excessive dipole moments as compared with experimental reference.
See also Ref.~\citenum{gillan2016perspective} and references therein.
Hence, for clathrate treated at the PBE level, methane interacts with water molecules that developed 
too large dipole moments.
The resulting error is not so obvious in the supermolecular interaction energy or in the lower-order terms,
but the overestimated many-body interactions accumulate in higher-body contributions, as seen in 
Table~\ref{four-body-rpa-energies}.
The SCAN and PBE0 functionals improve the description of water properties over PBE.\cite{chen2017ab,adamo1999accurate} 
Consequently, they also reduce the many-body errors in clathrate both at the self-consistent DFT level
and when used as an input for RPA.
Within RPA, the problem is also significantly reduced by the addition of the RSE term:
the fourth-order MBE for RPA(PBE)+RSE and RPA(PBE0)+RSE are converged to within 0.1 and 0.01~kcal/mol,
respectively.

Finally, let us discuss the accuracy of RPA for the subsets of four-body terms 
containing fragments with different number of hydrogen bonds, {\it i.e.}, 
subsets 4b 2hb, 4b 1hb, and 4b 0hb (Table~\ref{rpa-all-subsets}).
On average, RPA(PBE)+RSE gives more repulsive contributions on all three subsets, 
even though the individual errors are again both positive and negative, see the SI.
The magnitude of the errors is consistently reduced upon going to RPA(PBE0)+RSE.
RPA(SCAN) and RPA(SCAN0) four-body terms are too attractive overall, which is 
primarily caused by too attractive energies in the 4b 2hb subset.
The errors for the other two subsets are marginal.

\begin{table}[hbt]
  \setlength{\tabcolsep}{5pt}
  \caption{Four-body and higher-order nonadditive interaction energies (kcal/mol).
    The four-body energies are sums over 1140 tetramers \ce{CH4(H2O)3}.}
  \label{four-body-rpa-energies}
  \begin{threeparttable}
  \begin{tabular}{lrrrr}
    \toprule
    & SCAN0   &  PBE0  &  SCAN & PBE \\
    \midrule
    \multicolumn{5}{c}{Four-body nonadditive contributions} \\
    \midrule
    RPA+RSE & 0.77 & 0.63 & 0.95 & 0.85 \\
    RPA & 0.50 & 1.32 & 0.35 & 3.08 \\
    DFT$^a$ & 1.35 & $-1.68$ & 1.94 & $-3.46$ \\
    CCSD(T) & \multicolumn{4}{c}{0.56} \\
    \midrule
    \multicolumn{5}{c}{$n>4$-body nonadditive contributions} \\
    \midrule
    RPA+RSE & 0.01 & 0.01 & $-0.01$ & $-0.09$ \\
    RPA     & 0.04 & $-0.26$ & 0.10 & $-0.92$ \\
    DFT$^a$ & $-0.10$ & 0.67 & $-0.27$ & 1.18 \\
    $E_\text{c}^b$ & 0.00 & 0.15 & $-0.05$ & 0.26 \\
    \bottomrule
  \end{tabular}
  \footnotesize{
  \begin{tablenotes}
  \item[$a$] Self-consistent field energies obtained with the AVQZ basis.
  \item[$b$] RPA correlation energy contribution extrapolated to the basis set limit (AVTZ$\rightarrow$AVQZ).
  \end{tablenotes}
  }
  \end{threeparttable}
\end{table}

\subsubsection{General remarks}  \label{general-remarks}

The accuracy of RPA for $n$-body clusters depends on
the amount of electron correlation in the MBE term.
When the contribution of the correlation energy is significant,
the results become strongly dependent on the exchange-correlation model.\cite{modrzejewski2020random}

Table~\ref{share-of-correlation} shows subsets of MBE contributions,
presented together with the fraction of the correlation energy.
The subsets with a particularly high correlation content
are dimers, trimers 1hb, and tetramers 2hb.
Indeed, those are the subsets where all RPA methods
show the most significant deviations from the reference
(Table~\ref{rpa-all-subsets}).
The remaining cases are subsystems where Hartree-Fock is already
qualitatively correct and the correlation part is moderate:
3b 0hb, 4b 1hb, and 4b 0hb.
For those systems, the best RPA methods, RPA(SCAN0) and RPA(PBE0)+RSE,
reach the accuracy of about 0.1~kcal/mol, which is
almost within the uncertainty of the benchmark.
For the systems with a moderate amount of correlation
RPA falls midway between MP2 and the reference,
see Table~\ref{rpa-all-subsets}.

\begin{table*}
  \setlength{\tabcolsep}{5pt}
        \caption{Contributions to the methane---water cage interaction energy divided into $n$-body terms ($n$b)
        and subsystems with $m$ hydrogen bonds ($m$hb). Energies are in kcal/mol.}
        \label{rpa-all-subsets}
\begin{tabular}{lrrrrrrrrr}
\toprule
Method  &  2b & 3b 1hb & 3b 0hb & 4b 2hb & 4b 1hb & 4b 0hb & Sum\\
\midrule
HF+CABS &  $3.827$ & $-1.483$ & $1.210$ & $0.164$ & $0.471$ & $-0.083$ & $4.107$\\
CCSD(T) &  $-6.310$ & $-0.398$ & $1.434$ & $0.056$ & $0.552$ & $-0.051$ & $-4.718$\\
\midrule
RPA(PBE)     & $ -3.874$ & $-1.849$ & $-0.160$ & $1.152$ & $1.551$ & $0.374$& $-2.806$\\
RPA(PBE)+RSE     & $ -6.699$ & $-0.470$ & $1.411$ & $0.132$ & $0.705$ & $0.017$& $-4.904$\\
RPA(PBE0)     & $ -4.142$ & $-1.083$ & $0.872$ & $0.416$ & $0.847$ & $0.055$& $-3.036$\\
RPA(PBE0)+RSE     & $ -5.409$ & $-0.608$ & $1.349$ & $0.090$ & $0.585$ & $-0.044$& $-4.038$\\
RPA(SCAN)     & $ -4.433$ & $-0.721$ & $1.316$ & $-0.113$ & $0.539$ & $-0.071$& $-3.484$\\
RPA(SCAN)+RSE     & $ -6.461$ & $-0.642$ & $1.049$ & $0.208$ & $0.725$ & $0.017$& $-5.104$\\
RPA(SCAN0)     & $ -4.350$ & $-0.668$ & $1.308$ & $-0.007$ & $0.561$ & $-0.052$& $-3.208$\\
RPA(SCAN0)+RSE     & $ -5.361$ & $-0.729$ & $1.127$ & $0.162$ & $0.634$ & $-0.026$& $-4.192$\\
RPA(HF)         & $-2.983$ & $-0.770$ & $1.303$ & ---  & --- & --- & --- \\
\bottomrule
\end{tabular}
\end{table*}

\begin{table}[tb]
\caption{Subsets of $n$-body clusters and the share of electron correlation in their MBE contributions.}
\label{share-of-correlation}
\begin{threeparttable}
\begin{tabular}{llrrr}
\toprule
Subset & $N^a$ & $E_\text{int}^b$ & $\left|\frac{E_\text{int,c}}{E_\text{int}}\right|^c$ & $\left|\frac{E_\text{disp}}{E_\text{int}}\right|^d$ \\
\midrule
2b & 20 & $-6.31$ & 1.6 & 1.8 \\
3b 1hb & 30 & $-0.40$ & 2.7 & 1.8 \\
3b 0hb & 160 & 1.43 & 0.2 & 0.3 \\
4b 2hb & 60 & 0.06 & 1.9 & --- \\
4b 1hb & 420 & 0.55 & 0.1 & ---\\
4b 0hb & 660 & $-0.05$ & 0.6 & --- \\
\bottomrule
\end{tabular}
  \footnotesize{
    \begin{tablenotes}
        \item[$a$] Number of clusters in each subset.
        \item[$b$] CCSD(T) reference energy (kcal/mol).
        \item[$c$] $E_\text{int,c}$ is the contribution of the coupled-cluster correlation
        to the (nonadditive) interaction energy.
      \item[$d$] $E_\text{disp}$ denotes the dispersion plus exchange-dispersion SAPT2+3(CCD) contribution
        (dimers) and the PBE0 coupled three-body dispersion energy (trimers).
    \end{tablenotes}
    }
\end{threeparttable}
\end{table}

\subsubsection{Incremental correction scheme}

Compared to the leading terms, {\it e.g.}, dimer interaction energies, the higher $n$-body 
nonadditivities usually require less sophisticated treatment of electron 
correlation.\cite{gora2011interaction,BeranPCCP2012:fragment}
This is beneficial for large clusters and molecular solids where the number of relevant
three- and four-body terms can be very large and their evaluation using a simpler scheme leads 
to a substantial reduction of computational requirements.
Still, the evaluation of a large number of individual contributions can lead to an accumulation of 
numerical errors due to finite precision arithmetic.\cite{richard2013understanding}
This problem can be avoided if the energy of the whole system is obtained with 
the simpler scheme within a single calculation. 
Moreover, within this calculation all MBE terms and individual contributions are included avoiding
the real-space cut-offs necessary when applying MBE to large systems.
What is then left is to replace the simpler level description of the most important interactions (short distance two-body 
and possibly also higher-body) with a more accurate one, such as CCSD(T).
The relevant contributions can be selected using standard distance, connectivity, or energy criteria.\cite{hermann2008ground,liu2017understanding,liu2020energy}
For molecular solids, this incremental correction or subtractive embedding scheme is frequently used
by combining the CCSD(T) treatment of dimers
with periodic calculation performed at the level of HF or force-field.\cite{hermann2008ground,BeranPCCP2012:fragment,cervinka2018abinitio}
For clathrate, one can note that the largest RPA errors occur for compact fragments, that is, dimers 
and also trimers and tetramers with hydrogen bonded water molecules (Table~\ref{rpa-all-subsets}).
Therefore, RPA is a suitable scheme for the incremental correction approach.
The accuracy of the predicted RPA interaction energies can be improved by 
replacing all or some of the less accurate terms with their CCSD(T) counterparts.
We denote this as the CC/RPA approach (due to its similarity methods developed in, 
{\it e.g.}, Refs.~\citenum{tuma2004hybrid} or \citenum{bludsky2008investigation}) and test its performance in the following.

Let us define the following subsets $\mathcal{L}_X$ of fragments:
\begin{equation}
\begin{aligned}
        \mathcal{L}_0 &= \varnothing \\
        \mathcal{L}_{\rm 2b} &= \{ \text{2b} \} \\
        \mathcal{L}_{\rm 3b\ 1hb} &= \{ \text{2b}, \text{3b 1hb} \} \\
        \mathcal{L}_{\rm 3b\ 0hb} &= \{ \text{2b}, \text{3b 1hb}, \text{3b 0hb} \} \\
                      & \vdots \\
        \mathcal{L}_\infty &= \{ \text{2b}, \text{3b}, \text{4b}, \ldots, \text{21b} \} \label{subsets}
\end{aligned}
\end{equation}
In the last set all the fragments are included. 
This is equivalent to the supermolecular calculation.
In practice, replacing RPA contributions with CCSD(T) for smaller fragments, up to $X$,
can be written in a compact way as
\begin{multline}
       E_\text{int}^{\text{CC/RPA},X} = \\
\sum_{\zeta \in \mathcal{L}_X} 
\left(E_\text{int}^\text{CCSD(T)}[\zeta]-E_\text{int}^\text{RPA}[\zeta]\right) + E_\text{int}^\text{RPA}[\mathcal{L}_\infty]
\label{eq:cc-rpa}
\end{multline}
This can be understood in a way that we start from an RPA calculation and perform MBE on the difference
between RPA and CCSD(T). 
Alternatively, one can view the scheme as combination of CCSD(T) many-body contributions
up to $X$ with RPA many-body terms beyond $X$:
\begin{multline}
      E_\text{int}^{\text{CC/RPA},X} = \\
\sum_{\zeta \in \mathcal{L}_X} E_\text{int}^\text{CCSD(T)}[\zeta] + \sum_{\zeta' \in \mathcal{L}_\infty \setminus \mathcal{L}_X} E_\text{int}^\text{RPA}[\zeta']
\label{eq:cc-rpa2}
\end{multline}

\begin{figure*}
  \includegraphics[width=0.8\textwidth]{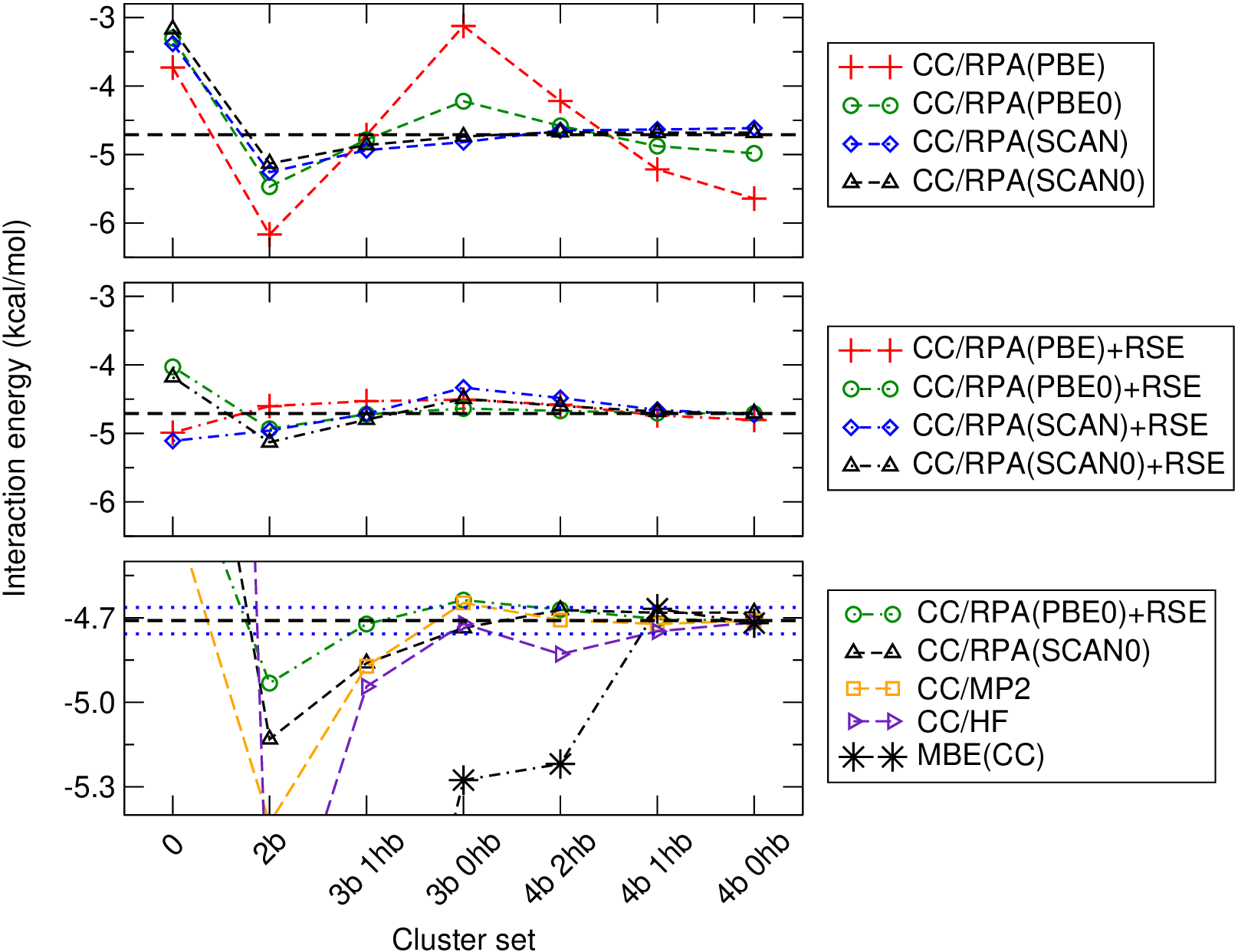}
  \caption{Methane---water cage interaction energy in the incremental correction scheme based on RPA
 (top panel) and  RPA+RSE (middle panel). 
  The bottom panel compares the two best CC/RPA approaches (CC/RPA(PBE0)+RSE and CC/RPA(SCAN0))
  to MBE at the CCSD(T) level and the correction scheme based on HF and MP2.
  On the $x$ axis, we give the index of cluster subset $\mathcal{L}_X$ up to which CCSD(T) 
 calculations are performed.
  In the bottom panel, the dashed line denotes the reference value of $-4.71$~kcal/mol 
  and the blue dotted lines are placed at values one percent above and below the reference.
 }
  \label{hybrid-mbe-rpa}
\end{figure*}

The incremental approach obviously works best when the RPA many-body contributions are close
to the CCSD(T) values.
Conversely, an inferior performance is expected when there is an error cancellation between different
n-body nonadditivities.
Specifically, as we deem the CCSD(T) MBE to be essentially converged using four-body terms, within 0.01~kcal/mol, 
there will be an error in the correction scheme if the five- and higher-body residual terms are 
significant. 
This can be seen from Eq.~\ref{eq:cc-rpa2}: the CC/RPA value for the complete set (up to 4b 0hb)
will be the CCSD(T) energy obtained using MBE plus the RPA higher-body residual terms.
The residual terms are particularly large for RPA based on the PBE and PBE0 functionals, being
$-0.92$ and $-0.26$~kcal/mol, respectively (Table~\ref{four-body-rpa-energies}).
These then cause the erroneous behavior of CC/RPA(PBE) and CC/RPA(PBE0) shown in Fig.~\ref{hybrid-mbe-rpa}.
The RSE corrections reduce the residual terms and error cancellations between $n$-body contributions, 
leading to improved performance of the correction scheme for both PBE- and PBE0-based RPA.
The error of CC/RPA(PBE0)+RSE is only 4.7~\% after replacing the two-body RPA terms
with CCSD(T).
Subsequent corrections keep the energy within 1.6~\% or 0.08~kcal/mol of the reference value.

The SCAN- and SCAN0-based RPA are much less prone to the error cancellations, even without RSE. 
Also note that the RSE terms degrade the accuracy of the three- and higher-order nonadditive energies.
As a consequence one can see that CC/RPA(SCAN) and CC/RPA(SCAN0) without RSE perform somewhat better 
than the variants with RSE.
Overall, the CC/RPA(SCAN0) scheme looks particularly promising. 
The error is around 3~\% or 0.15~kcal/mol when terms up to 3b~1hb are treated with CCSD(T).
When the 3b~0hb terms are added, the predicted interaction energy gets 
within 1~\% or 0.05~kcal/mol of the reference.

We now compare the CC/RPA scheme to similar approaches: CC/HF and CC/MP2 correction schemes 
defined analogously to Eq.~\ref{eq:cc-rpa} and a standard MBE based on CCSD(T) where only contributions
up to a given order are accounted for.
In the last approach, denoted MBE(CC) in Fig.~\ref{hybrid-mbe-rpa},
one needs to include the terms up to 4b~1hb to reduce the error below 10~\%.
The error originates mostly from the neglect of the HF contribution,
which dominates the terms above and including 3b~0hb.
As shown in Fig.~\ref{hybrid-mbe-rpa}, the results are improved in the CC/HF correction scheme.
However, the CC/HF results are not completely satisfactory as the error is still around 0.1~kcal/mol 
even when the 4b 2hb clusters are treated at the CCSD(T) level.

Clearly, a correction scheme with a correlated supermolecular reference, as done in CC/RPA, would reduce the need 
to correct the four-body contributions.
For clathrate, MP2 is the only M{\o}ller-Plesset perturbation-theory method for which the supermolecular calculation
is feasible with a large basis set.
(See the Supporting Information for the comparison of CPU times, {\it e.g.},
between MP2 and MP3.)
The CC/MP2 reduces the errors compared to CC/HF and the differences to the reference are negligible (below 0.01~kcal/mol) 
once the 4b 2hb contributions are corrected.
Compared to CC/RPA(PBE0)+RSE, CC/MP2 has larger errors when the 3b 1hb and 3b 0hb subsets are not corrected
since MP2 lacks the description of three-body dispersion.
When 4b 2hb and larger clusters are corrected, CC/RPA(PBE0)+RSE, CC/RPA(SCAN0), and CC/MP2 show 
comparable results. 
Finally, we note that with tight precision settings the computational cost of RPA grows less steeply with system size 
than that of MP2.
This offers an advantage for the CC/RPA scheme for systems such as molecular solids where converging
MP2 with k-points is very difficult.\cite{delben2013electron,liao2016finite}

\section{Conclusions}
We have reported the most accurate to date estimate of the interaction energy
of the methane molecule in a dodecahedral water cage, a finite model
of the many-body noncovalent interactions present in molecular solids.
This reference energy is obtained as a sum of many-body contributions at the complete basis set limit, 
computed through the four-body terms at the frozen-core CCSD(T) level,
with the remaining, almost negligible higher-order contributions estimated
at the Hartree-Fock level.
We have analyzed the convergence of each MBE term with the basis-set cardinal number
and compared the canonical and F12 variants of CCSD(T) for additional validation.
We used the benchmark data to analyze the performance of approximate methods:
M{\o}ller-Plesset perturbation theory, density-functional models, and, which was
the primary focus, the post-Kohn-Sham random-phase approximation.

Among tested methods, MP3 yields near-benchmark accuracy for the nonadditive
contributions of trimers and tetramers, but is too shallow by ca.~1~kcal/mol
for dimers.
The challenging dimer contribution is recovered within 0.05~kcal/mol by MP4.
All tested semilocal and hybrid DFT approximations,
PBE, SCAN, PBE0, and SCAN0, behave erratically,
with three- and four-body nonadditive terms having neither the correct sign
nor magnitude.
The post-Kohn-Sham random-phase approximation, which uses the approximate
Kohn-Sham eigenstates as its input, does not suffer from those issues,
but we find that it is crucial to apply it with the renormalized singles
correction energy.

The RSE term is important in all orders of the many-body expansion.
For the \ce{CH4\bond{...}H2O} dimers, RSE corrects the well-known
underestimation of the two-body RPA interaction energy.
This holds for all tested orbitals: pure and hybrid GGAs and meta-GGAs 
as well as the orbitals generated with optimized effective
potentials (OEPX and OEP2-sc).

Remarkably, the inclusion of singles can also correct wrong signs of three-
and four-body nonadditive energies as well as mitigate excessive higher-order contributions
to the many-body expansion.
This point is particularly relevant for RPA based on the PBE and PBE0 functionals.
For the methane clathrate, a nonpolar system, the sum of $n>4$-body terms is expected
to be negligible, but it is as large as $-0.92$ and $-0.26$~kcal/mol
for RPA(PBE) and RPA(PBE0), respectively.
This erratic behavior is largely corrected by the RSE term.

RPA based on SCAN and SCAN0 eigenstates shows a different behavior compared
to RPA based on PBE or PBE0.
The nonadditive energies of RPA(SCAN) and
RPA(SCAN0) are close to those of RPA(HF).
For this reason, the RSE contribution to the $n\ge 3$-body nonadditivities
is smaller than for the PBE-based variants and does not introduce
a qualitative change.

Finally, the errors in the RPA interaction energy stem mainly from short-range compact clusters, 
{\it i.e.}, dimers and trimers with hydrogen bonded water molecules.
RPA is therefore suitable for the incremental correction scheme, 
where CCSD(T) is used for the problematic compact clusters and RPA for the remainder via a supermolecular calculation.
For the clathrate considered here, treating the twenty dimers and thirty hydrogen-bonded trimers at the CCSD(T)
level and using RPA(PBE0)+RSE for the rest, the incrementally corrected interaction energy
lies within 0.08~kcal/mol or 1.6~{\%} of the reference.
This result makes RPA/CC promising for large clusters or molecular solids
where the correction scheme avoids the explicit evaluation of the large number of distant two- and three-body
contributions and higher-body terms.\cite{BeranPCCP2012:fragment}

\begin{acknowledgement}
  This work was supported by the European Research Council (ERC)
  under the European Union's Horizon 2020 research and innovation 
  program (grant agreement No 759721) and by the Primus programme 
  of the Charles University.
  S.Ś. is grateful to the Polish National Science Center for the partial
  financial support under Grant No. 2016/21/D/ST4/00903.
  This research was supported in part by PLGrid Infrastructure.
We are also grateful for computational resources provided by 
the IT4Innovations National Supercomputing Center  (LM2015070), 
CESNET (LM2015042), CERIT Scientific Cloud (LM2015085), and 
e-Infrastruktura~CZ (e-INFRA LM2018140).
  The authors thank Prof.~Grzegorz Chałasiński for useful 
discussions related to this work.
We thank Aleksandra Tucholska for help with the table of contents graphics.

\end{acknowledgement}

\begin{suppinfo}
Spreadsheets with raw numerical data and computational details,
geometries in a form of xyz files. 
The outputs of calculations and processing script are available 
at {\tt https://github.com/klimes/Clathrate\_cluster\_data}
and in the Zenodo repository at https://doi.org/10.5281/zenodo.4429677
\end{suppinfo}

{\footnotesize
\bibliography{biblio}}

\providecommand{\latin}[1]{#1}
\providecommand*\mcitethebibliography{\thebibliography}
\csname @ifundefined\endcsname{endmcitethebibliography}
  {\let\endmcitethebibliography\endthebibliography}{}
\begin{mcitethebibliography}{105}
\providecommand*\natexlab[1]{#1}
\providecommand*\mciteSetBstSublistMode[1]{}
\providecommand*\mciteSetBstMaxWidthForm[2]{}
\providecommand*\mciteBstWouldAddEndPuncttrue
  {\def\EndOfBibitem{\unskip.}}
\providecommand*\mciteBstWouldAddEndPunctfalse
  {\let\EndOfBibitem\relax}
\providecommand*\mciteSetBstMidEndSepPunct[3]{}
\providecommand*\mciteSetBstSublistLabelBeginEnd[3]{}
\providecommand*\EndOfBibitem{}
\mciteSetBstSublistMode{f}
\mciteSetBstMaxWidthForm{subitem}{(\alph{mcitesubitemcount})}
\mciteSetBstSublistLabelBeginEnd
  {\mcitemaxwidthsubitemform\space}
  {\relax}
  {\relax}

\bibitem[Jure\v{c}ka \latin{et~al.}(2006)Jure\v{c}ka, \v{S}poner,
  \v{C}ern\'{y}, and Hobza]{jurecka2006benchmark}
Jure\v{c}ka,~P.; \v{S}poner,~J.; \v{C}ern\'{y},~J.; Hobza,~P. {Benchmark
  database of accurate (MP2 and CCSD(T) complete basis set limit) interaction
  energies of small model complexes, DNA base pairs, and amino acid pairs}.
  \emph{Phys. Chem. Chem. Phys.} \textbf{2006}, \emph{8}, 1985--1993\relax
\mciteBstWouldAddEndPuncttrue
\mciteSetBstMidEndSepPunct{\mcitedefaultmidpunct}
{\mcitedefaultendpunct}{\mcitedefaultseppunct}\relax
\EndOfBibitem
\bibitem[Peverati and Truhlar(2014)Peverati, and Truhlar]{peverati2014quest}
Peverati,~R.; Truhlar,~D.~G. Quest for a universal density functional: the
  accuracy of density functionals across a broad spectrum of databases in
  chemistry and physics. \emph{Phil. Trans. R. Soc. A} \textbf{2014},
  \emph{372}, 20120476\relax
\mciteBstWouldAddEndPuncttrue
\mciteSetBstMidEndSepPunct{\mcitedefaultmidpunct}
{\mcitedefaultendpunct}{\mcitedefaultseppunct}\relax
\EndOfBibitem
\bibitem[Taylor \latin{et~al.}(2016)Taylor, Ángyán, Galli, Zhang, Gygi,
  Hirao, Song, Rahul, Anatole~von Lilienfeld, Podeszwa, Bulik, Henderson,
  Scuseria, Toulouse, Peverati, Truhlar, and Szalewicz]{taylor2016blind}
Taylor,~D.~E.; Ángyán,~J.~G.; Galli,~G.; Zhang,~C.; Gygi,~F.; Hirao,~K.;
  Song,~J.~W.; Rahul,~K.; Anatole~von Lilienfeld,~O.; Podeszwa,~R.;
  Bulik,~I.~W.; Henderson,~T.~M.; Scuseria,~G.~E.; Toulouse,~J.; Peverati,~R.;
  Truhlar,~D.~G.; Szalewicz,~K. Blind test of density-functional-based methods
  on intermolecular interaction energies. \emph{J. Chem. Phys.} \textbf{2016},
  \emph{145}, 124105\relax
\mciteBstWouldAddEndPuncttrue
\mciteSetBstMidEndSepPunct{\mcitedefaultmidpunct}
{\mcitedefaultendpunct}{\mcitedefaultseppunct}\relax
\EndOfBibitem
\bibitem[Goerigk \latin{et~al.}(2017)Goerigk, Hansen, Bauer, Ehrlich, Najibi,
  and Grimme]{goerigk2017a}
Goerigk,~L.; Hansen,~A.; Bauer,~C.; Ehrlich,~S.; Najibi,~A.; Grimme,~S. A look
  at the density functional theory zoo with the advanced GMTKN55 database for
  general main group thermochemistry{,} kinetics and noncovalent interactions.
  \emph{Phys. Chem. Chem. Phys.} \textbf{2017}, \emph{19}, 32184--32215\relax
\mciteBstWouldAddEndPuncttrue
\mciteSetBstMidEndSepPunct{\mcitedefaultmidpunct}
{\mcitedefaultendpunct}{\mcitedefaultseppunct}\relax
\EndOfBibitem
\bibitem[\v{R}ez\'{a}\v{c} \latin{et~al.}(2015)\v{R}ez\'{a}\v{c}, Huang, Hobza,
  and Beran]{rezac2015benchmark}
\v{R}ez\'{a}\v{c},~J.; Huang,~Y.; Hobza,~P.; Beran,~G. J.~O. Benchmark
  calculations of three-body intermolecular interactions and the performance of
  low-cost electronic structure methods. \emph{J. Chem. Theory Comput.}
  \textbf{2015}, \emph{11}, 3065\relax
\mciteBstWouldAddEndPuncttrue
\mciteSetBstMidEndSepPunct{\mcitedefaultmidpunct}
{\mcitedefaultendpunct}{\mcitedefaultseppunct}\relax
\EndOfBibitem
\bibitem[Huang and Beran(2015)Huang, and Beran]{huang2015reliable}
Huang,~Y.; Beran,~G. J.~O. Reliable prediction of three-body intermolecular
  interactions using dispersion-corrected second-order Møller-Plesset
  perturbation theory. \emph{J. Chem. Phys.} \textbf{2015}, \emph{143},
  044113\relax
\mciteBstWouldAddEndPuncttrue
\mciteSetBstMidEndSepPunct{\mcitedefaultmidpunct}
{\mcitedefaultendpunct}{\mcitedefaultseppunct}\relax
\EndOfBibitem
\bibitem[Jankiewicz \latin{et~al.}(2018)Jankiewicz, Podeszwa, and
  Witek]{jankiewicz2018dispersion}
Jankiewicz,~W.; Podeszwa,~R.; Witek,~H.~A. Dispersion-Corrected DFT Struggles
  with Predicting Three-Body Interaction Energies. \emph{J. Chem. Theory
  Comput.} \textbf{2018}, \emph{14}, 5079--5089\relax
\mciteBstWouldAddEndPuncttrue
\mciteSetBstMidEndSepPunct{\mcitedefaultmidpunct}
{\mcitedefaultendpunct}{\mcitedefaultseppunct}\relax
\EndOfBibitem
\bibitem[Lotrich and Szalewicz(1997)Lotrich, and Szalewicz]{lotrich1997three}
Lotrich,~V.~F.; Szalewicz,~K. Three-body contribution to binding energy of
  solid argon and analysis of crystal structure. \emph{Phys. Rev. Lett.}
  \textbf{1997}, \emph{79}, 1301\relax
\mciteBstWouldAddEndPuncttrue
\mciteSetBstMidEndSepPunct{\mcitedefaultmidpunct}
{\mcitedefaultendpunct}{\mcitedefaultseppunct}\relax
\EndOfBibitem
\bibitem[Yang \latin{et~al.}(2014)Yang, Hu, Usvyat, Matthews, Sch{\"u}tz, and
  Chan]{yang2014determination}
Yang,~J.; Hu,~W.; Usvyat,~D.; Matthews,~D.; Sch{\"u}tz,~M.; Chan,~G. K.-L. Ab
  initio determination of the crystalline benzene lattice energy to
  sub-kilojoule/mole accuracy. \emph{Science} \textbf{2014}, \emph{345},
  640--643\relax
\mciteBstWouldAddEndPuncttrue
\mciteSetBstMidEndSepPunct{\mcitedefaultmidpunct}
{\mcitedefaultendpunct}{\mcitedefaultseppunct}\relax
\EndOfBibitem
\bibitem[Kennedy \latin{et~al.}(2014)Kennedy, McDonald, DePrince, Marshall,
  Podeszwa, and Sherrill]{kennedy2014resolving}
Kennedy,~M.~R.; McDonald,~A.~R.; DePrince,~A.~E.; Marshall,~M.~S.;
  Podeszwa,~R.; Sherrill,~C.~D. {Communication: Resolving the three-body
  contribution to the lattice energy of crystalline benzene: Benchmark results
  from coupled-cluster theory}. \emph{J. Chem. Phys.} \textbf{2014},
  \emph{140}, 121104\relax
\mciteBstWouldAddEndPuncttrue
\mciteSetBstMidEndSepPunct{\mcitedefaultmidpunct}
{\mcitedefaultendpunct}{\mcitedefaultseppunct}\relax
\EndOfBibitem
\bibitem[Gillan(2014)]{gillan2014many}
Gillan,~M. Many-body exchange-overlap interactions in rare gases and water.
  \emph{J. Chem. Phys.} \textbf{2014}, \emph{141}, 224106\relax
\mciteBstWouldAddEndPuncttrue
\mciteSetBstMidEndSepPunct{\mcitedefaultmidpunct}
{\mcitedefaultendpunct}{\mcitedefaultseppunct}\relax
\EndOfBibitem
\bibitem[Hapka \latin{et~al.}(2017)Hapka, Rajchel, Modrzejewski, Schaeffer,
  Chalasinski, and Szczesniak]{hapka2017nature}
Hapka,~M.; Rajchel,~L.; Modrzejewski,~M.; Schaeffer,~R.; Chalasinski,~G.;
  Szczesniak,~M.~M. The nature of three-body interactions in DFT: Exchange and
  polarization effects. \emph{J. Chem. Phys.} \textbf{2017}, \emph{147},
  084106\relax
\mciteBstWouldAddEndPuncttrue
\mciteSetBstMidEndSepPunct{\mcitedefaultmidpunct}
{\mcitedefaultendpunct}{\mcitedefaultseppunct}\relax
\EndOfBibitem
\bibitem[Eshuis \latin{et~al.}(2012)Eshuis, Bates, and
  Furche]{eshuis2012electron}
Eshuis,~H.; Bates,~J.~E.; Furche,~F. Electron correlation methods based on the
  random phase approximation. \emph{Theor. Chem. Acc.} \textbf{2012},
  \emph{131}, 1084\relax
\mciteBstWouldAddEndPuncttrue
\mciteSetBstMidEndSepPunct{\mcitedefaultmidpunct}
{\mcitedefaultendpunct}{\mcitedefaultseppunct}\relax
\EndOfBibitem
\bibitem[Dobson(2012)]{dobson2012dispersion}
Dobson,~J.~F. In \emph{{Fundamentals of Time-Dependent Density Functional
  Theory}}; Marques,~M.~A., Maitra,~N.~T., Nogueira,~F.~M., Gross,~E.,
  Rubio,~A., Eds.; Springer: Berlin, Heidelberg, 2012; pp 417--441\relax
\mciteBstWouldAddEndPuncttrue
\mciteSetBstMidEndSepPunct{\mcitedefaultmidpunct}
{\mcitedefaultendpunct}{\mcitedefaultseppunct}\relax
\EndOfBibitem
\bibitem[Dobson(2014)]{dobson2014beyond}
Dobson,~J.~F. Beyond pairwise additivity in London dispersion interactions.
  \emph{Int. J. Quant. Chem.} \textbf{2014}, \emph{114}, 1157\relax
\mciteBstWouldAddEndPuncttrue
\mciteSetBstMidEndSepPunct{\mcitedefaultmidpunct}
{\mcitedefaultendpunct}{\mcitedefaultseppunct}\relax
\EndOfBibitem
\bibitem[Hermann \latin{et~al.}(2017)Hermann, DiStasio~Jr, and
  Tkatchenko]{hermann2017first}
Hermann,~J.; DiStasio~Jr,~R.~A.; Tkatchenko,~A. First-Principles Models for van
  der Waals Interactions in Molecules and Materials: Concepts, Theory, and
  Applications. \emph{Chem. Rev.} \textbf{2017}, \emph{117}, 4714\relax
\mciteBstWouldAddEndPuncttrue
\mciteSetBstMidEndSepPunct{\mcitedefaultmidpunct}
{\mcitedefaultendpunct}{\mcitedefaultseppunct}\relax
\EndOfBibitem
\bibitem[Richard \latin{et~al.}(2014)Richard, Lao, and
  Herbert]{richard2014aiming}
Richard,~R.~M.; Lao,~K.~U.; Herbert,~J.~M. Aiming for Benchmark Accuracy with
  the Many-Body Expansion. \emph{Acc. Chem. Res.} \textbf{2014}, \emph{47},
  2828--2836\relax
\mciteBstWouldAddEndPuncttrue
\mciteSetBstMidEndSepPunct{\mcitedefaultmidpunct}
{\mcitedefaultendpunct}{\mcitedefaultseppunct}\relax
\EndOfBibitem
\bibitem[Eshuis \latin{et~al.}(2010)Eshuis, Yarkony, and
  Furche]{eshuis2010fast}
Eshuis,~H.; Yarkony,~J.; Furche,~F. Fast computation of molecular random phase
  approximation correlation energies using resolution of the identity and
  imaginary frequency integration. \emph{J. Chem. Phys.} \textbf{2010},
  \emph{132}, 234114\relax
\mciteBstWouldAddEndPuncttrue
\mciteSetBstMidEndSepPunct{\mcitedefaultmidpunct}
{\mcitedefaultendpunct}{\mcitedefaultseppunct}\relax
\EndOfBibitem
\bibitem[Del~Ben \latin{et~al.}(2013)Del~Ben, Hutter, and
  VandeVondele]{delben2013electron}
Del~Ben,~M.; Hutter,~J.; VandeVondele,~J. Electron Correlation in the Condensed
  Phase from a Resolution of Identity Approach Based on the Gaussian and Plane
  Waves Scheme. \emph{J. Chem. Theory Comput.} \textbf{2013}, \emph{9},
  2654--2671\relax
\mciteBstWouldAddEndPuncttrue
\mciteSetBstMidEndSepPunct{\mcitedefaultmidpunct}
{\mcitedefaultendpunct}{\mcitedefaultseppunct}\relax
\EndOfBibitem
\bibitem[Kaltak \latin{et~al.}(2014)Kaltak, Klimes, and Kresse]{kaltak2014low}
Kaltak,~M.; Klimes,~J.; Kresse,~G. {Low scaling algorithms for the random phase
  approximation: Imaginary time and laplace transformations}. \emph{J. Chem.
  Theory Comput.} \textbf{2014}, \emph{10}, 2498\relax
\mciteBstWouldAddEndPuncttrue
\mciteSetBstMidEndSepPunct{\mcitedefaultmidpunct}
{\mcitedefaultendpunct}{\mcitedefaultseppunct}\relax
\EndOfBibitem
\bibitem[Kaltak \latin{et~al.}(2014)Kaltak, Klimes, and
  Kresse]{kaltak2014cubic}
Kaltak,~M.; Klimes,~J.; Kresse,~G. {Cubic scaling algorithm for the random
  phase approximation: Self-interstitials and vacancies in Si}. \emph{Phys.
  Rev. B} \textbf{2014}, \emph{90}, 054115\relax
\mciteBstWouldAddEndPuncttrue
\mciteSetBstMidEndSepPunct{\mcitedefaultmidpunct}
{\mcitedefaultendpunct}{\mcitedefaultseppunct}\relax
\EndOfBibitem
\bibitem[Schurkus and Ochsenfeld(2016)Schurkus, and
  Ochsenfeld]{schurkus2016effective}
Schurkus,~H.~F.; Ochsenfeld,~C. {Communication: An effective linear-scaling
  atomic-orbital reformulation of the random-phase approximation using a
  contracted double-Laplace transformation}. \emph{J. Chem. Phys.}
  \textbf{2016}, \emph{144}, 031101\relax
\mciteBstWouldAddEndPuncttrue
\mciteSetBstMidEndSepPunct{\mcitedefaultmidpunct}
{\mcitedefaultendpunct}{\mcitedefaultseppunct}\relax
\EndOfBibitem
\bibitem[Wilhelm \latin{et~al.}(2016)Wilhelm, Seewald, Del~Ben, and
  Hutter]{wilhelm2016large}
Wilhelm,~J.; Seewald,~P.; Del~Ben,~M.; Hutter,~J. {Large-scale cubic-scaling
  random phase approximation correlation energy calculations using a Gaussian
  basis}. \emph{J. Chem. Theory Comput.} \textbf{2016}, \emph{12}, 5851\relax
\mciteBstWouldAddEndPuncttrue
\mciteSetBstMidEndSepPunct{\mcitedefaultmidpunct}
{\mcitedefaultendpunct}{\mcitedefaultseppunct}\relax
\EndOfBibitem
\bibitem[Klime{\v s}(2016)]{klimes2016lattice}
Klime{\v s},~J. Lattice energies of molecular solids from the random phase
  approximation with singles corrections. \emph{J. Chem. Phys.} \textbf{2016},
  \emph{145}, 094506\relax
\mciteBstWouldAddEndPuncttrue
\mciteSetBstMidEndSepPunct{\mcitedefaultmidpunct}
{\mcitedefaultendpunct}{\mcitedefaultseppunct}\relax
\EndOfBibitem
\bibitem[Modrzejewski \latin{et~al.}(2020)Modrzejewski, Yourdkhani, and
  Klime\v{s}]{modrzejewski2020random}
Modrzejewski,~M.; Yourdkhani,~S.; Klime\v{s},~J. {Random Phase Approximation
  Applied to Many-Body Noncovalent Systems}. \emph{J. Chem. Theory Comput.}
  \textbf{2020}, \emph{16}, 427--442\relax
\mciteBstWouldAddEndPuncttrue
\mciteSetBstMidEndSepPunct{\mcitedefaultmidpunct}
{\mcitedefaultendpunct}{\mcitedefaultseppunct}\relax
\EndOfBibitem
\bibitem[Ambrosetti \latin{et~al.}(2014)Ambrosetti, Reilly, DiStasio~Jr, and
  Tkatchenko]{ambrosetti2014long}
Ambrosetti,~A.; Reilly,~A.~M.; DiStasio~Jr,~R.~A.; Tkatchenko,~A. Long-range
  correlation energy calculated from coupled atomic response functions.
  \emph{J. Chem. Phys.} \textbf{2014}, \emph{140}, 18A508\relax
\mciteBstWouldAddEndPuncttrue
\mciteSetBstMidEndSepPunct{\mcitedefaultmidpunct}
{\mcitedefaultendpunct}{\mcitedefaultseppunct}\relax
\EndOfBibitem
\bibitem[Kim \latin{et~al.}(2020)Kim, Kim, Gould, Lee, Leb{\`e}gue, and
  Kim]{kim2020umbd}
Kim,~M.; Kim,~W.~J.; Gould,~T.; Lee,~E.~K.; Leb{\`e}gue,~S.; Kim,~H. {uMBD: A
  Materials-Ready Dispersion Correction That Uniformly Treats Metallic, Ionic,
  and van der Waals Bonding}. \emph{J. Am. Chem. Soc.} \textbf{2020},
  \emph{142}, 2346--2354\relax
\mciteBstWouldAddEndPuncttrue
\mciteSetBstMidEndSepPunct{\mcitedefaultmidpunct}
{\mcitedefaultendpunct}{\mcitedefaultseppunct}\relax
\EndOfBibitem
\bibitem[He{\ss}elmann(2018)]{hesselmann2018correlation}
He{\ss}elmann,~A. Correlation effects and many-body interactions in water
  clusters. \emph{Beilstein J. Org. Chem.} \textbf{2018}, \emph{14},
  979--991\relax
\mciteBstWouldAddEndPuncttrue
\mciteSetBstMidEndSepPunct{\mcitedefaultmidpunct}
{\mcitedefaultendpunct}{\mcitedefaultseppunct}\relax
\EndOfBibitem
\bibitem[Deible \latin{et~al.}(2014)Deible, Tuguldur, and
  Jordan]{deible2014theoretical}
Deible,~M.~J.; Tuguldur,~O.; Jordan,~K.~D. Theoretical Study of the Binding
  Energy of a Methane Molecule in a (H$_{2}$O)$_{20}$ Dodecahedral Cage.
  \emph{J. Phys. Chem. B} \textbf{2014}, \emph{118}, 8257--8263\relax
\mciteBstWouldAddEndPuncttrue
\mciteSetBstMidEndSepPunct{\mcitedefaultmidpunct}
{\mcitedefaultendpunct}{\mcitedefaultseppunct}\relax
\EndOfBibitem
\bibitem[Gillan \latin{et~al.}(2015)Gillan, Alfè, and Manby]{gillan2015energy}
Gillan,~M.~J.; Alfè,~D.; Manby,~F.~R. Energy benchmarks for methane-water
  systems from quantum Monte Carlo and second-order Møller-Plesset
  calculations. \emph{J. Chem. Phys.} \textbf{2015}, \emph{143}, 102812\relax
\mciteBstWouldAddEndPuncttrue
\mciteSetBstMidEndSepPunct{\mcitedefaultmidpunct}
{\mcitedefaultendpunct}{\mcitedefaultseppunct}\relax
\EndOfBibitem
\bibitem[Cox \latin{et~al.}(2014)Cox, Towler, Alf{\`e}, and
  Michaelides]{cox2014benchmarking}
Cox,~S.~J.; Towler,~M.~D.; Alf{\`e},~D.; Michaelides,~A. Benchmarking the
  performance of density functional theory and point charge force fields in
  their description of sI methane hydrate against diffusion Monte Carlo.
  \emph{J. Chem. Phys.} \textbf{2014}, \emph{140}, 174703\relax
\mciteBstWouldAddEndPuncttrue
\mciteSetBstMidEndSepPunct{\mcitedefaultmidpunct}
{\mcitedefaultendpunct}{\mcitedefaultseppunct}\relax
\EndOfBibitem
\bibitem[Perdew \latin{et~al.}(1996)Perdew, Burke, and
  Ernzerhof]{perdew1996generalized}
Perdew,~J.; Burke,~K.; Ernzerhof,~M. {Generalized Gradient Approximation Made
  Simple}. \emph{Phys. Rev. Lett.} \textbf{1996}, \emph{77}, 3865--3868\relax
\mciteBstWouldAddEndPuncttrue
\mciteSetBstMidEndSepPunct{\mcitedefaultmidpunct}
{\mcitedefaultendpunct}{\mcitedefaultseppunct}\relax
\EndOfBibitem
\bibitem[Adamo and Barone(1999)Adamo, and Barone]{adamo1999toward}
Adamo,~C.; Barone,~V. {Toward reliable density functional methods without
  adjustable parameters: The PBE0 model}. \emph{J. Chem. Phys.} \textbf{1999},
  \emph{110}, 6158--6170\relax
\mciteBstWouldAddEndPuncttrue
\mciteSetBstMidEndSepPunct{\mcitedefaultmidpunct}
{\mcitedefaultendpunct}{\mcitedefaultseppunct}\relax
\EndOfBibitem
\bibitem[Sun \latin{et~al.}(2015)Sun, Ruzsinszky, and Perdew]{sun2015strongly}
Sun,~J.; Ruzsinszky,~A.; Perdew,~J.~P. {Strongly Constrained and Appropriately
  Normed Semilocal Density Functional}. \emph{Phys. Rev. Lett.} \textbf{2015},
  \emph{115}, 036402\relax
\mciteBstWouldAddEndPuncttrue
\mciteSetBstMidEndSepPunct{\mcitedefaultmidpunct}
{\mcitedefaultendpunct}{\mcitedefaultseppunct}\relax
\EndOfBibitem
\bibitem[Hui and Chai(2016)Hui, and Chai]{hui2016scan}
Hui,~K.; Chai,~J.-D. {SCAN-based hybrid and double-hybrid density functionals
  from models without fitted parameters}. \emph{J. Chem. Phys.} \textbf{2016},
  \emph{144}, 044114\relax
\mciteBstWouldAddEndPuncttrue
\mciteSetBstMidEndSepPunct{\mcitedefaultmidpunct}
{\mcitedefaultendpunct}{\mcitedefaultseppunct}\relax
\EndOfBibitem
\bibitem[Perdew \latin{et~al.}(2005)Perdew, Ruzsinszky, Tao, Staroverov,
  Scuseria, and Csonka]{perdew2005prescription}
Perdew,~J.; Ruzsinszky,~A.; Tao,~J.; Staroverov,~V.; Scuseria,~G.; Csonka,~G.
  {Prescription for the design and selection of density functional
  approximations: More constraint satisfaction with fewer fits}. \emph{J. Chem.
  Phys.} \textbf{2005}, \emph{123}, 062201\relax
\mciteBstWouldAddEndPuncttrue
\mciteSetBstMidEndSepPunct{\mcitedefaultmidpunct}
{\mcitedefaultendpunct}{\mcitedefaultseppunct}\relax
\EndOfBibitem
\bibitem[Klimes \latin{et~al.}(2015)Klimes, Kaltak, Maggio, and
  Kresse]{klimes2015singles}
Klimes,~J.; Kaltak,~M.; Maggio,~E.; Kresse,~G. Singles correlation energy
  contributions in solids. \emph{J. Chem. Phys.} \textbf{2015}, \emph{143},
  102816\relax
\mciteBstWouldAddEndPuncttrue
\mciteSetBstMidEndSepPunct{\mcitedefaultmidpunct}
{\mcitedefaultendpunct}{\mcitedefaultseppunct}\relax
\EndOfBibitem
\bibitem[Ren \latin{et~al.}(2013)Ren, Rinke, Scuseria, and
  Scheffler]{ren2013renormalized}
Ren,~X.; Rinke,~P.; Scuseria,~G.~E.; Scheffler,~M. {Renormalized second-order
  perturbation theory for the electron correlation energy: Concept,
  implementation, and benchmarks}. \emph{Phys. Rev. B} \textbf{2013},
  \emph{88}, 035120\relax
\mciteBstWouldAddEndPuncttrue
\mciteSetBstMidEndSepPunct{\mcitedefaultmidpunct}
{\mcitedefaultendpunct}{\mcitedefaultseppunct}\relax
\EndOfBibitem
\bibitem[Werner \latin{et~al.}(2012)Werner, Knowles, Knizia, Manby, and
  Sch{\"u}tz]{werner2012molpro}
Werner,~H.-J.; Knowles,~P.~J.; Knizia,~G.; Manby,~F.~R.; Sch{\"u}tz,~M. Molpro:
  a general-purpose quantum chemistry program package. \emph{WIREs Comput. Mol.
  Sci.} \textbf{2012}, \emph{2}, 242--253\relax
\mciteBstWouldAddEndPuncttrue
\mciteSetBstMidEndSepPunct{\mcitedefaultmidpunct}
{\mcitedefaultendpunct}{\mcitedefaultseppunct}\relax
\EndOfBibitem
\bibitem[Kendall \latin{et~al.}(1992)Kendall, Dunning~Jr, and
  Harrison]{kendall1992electron}
Kendall,~R.~A.; Dunning~Jr,~T.~H.; Harrison,~R.~J. Electron affinities of the
  first-row atoms revisited. Systematic basis sets and wave functions. \emph{J.
  Chem. Phys.} \textbf{1992}, \emph{96}, 6796--6806\relax
\mciteBstWouldAddEndPuncttrue
\mciteSetBstMidEndSepPunct{\mcitedefaultmidpunct}
{\mcitedefaultendpunct}{\mcitedefaultseppunct}\relax
\EndOfBibitem
\bibitem[Schuchardt \latin{et~al.}(2007)Schuchardt, Didier, Elsethagen, Sun,
  Gurumoorthi, Chase, Li, and Windus]{schuchardt2007basis}
Schuchardt,~K.~L.; Didier,~B.~T.; Elsethagen,~T.; Sun,~L.; Gurumoorthi,~V.;
  Chase,~J.; Li,~J.; Windus,~T.~L. {Basis Set Exchange: A Community Database
  for Computational Sciences}. \emph{J. Chem. Inf. Model.} \textbf{2007},
  \emph{47}, 1045--1052\relax
\mciteBstWouldAddEndPuncttrue
\mciteSetBstMidEndSepPunct{\mcitedefaultmidpunct}
{\mcitedefaultendpunct}{\mcitedefaultseppunct}\relax
\EndOfBibitem
\bibitem[Halkier \latin{et~al.}(1999)Halkier, Klopper, Helgaker, Jorgensen, and
  Taylor]{halkier1999basis}
Halkier,~A.; Klopper,~W.; Helgaker,~T.; Jorgensen,~P.; Taylor,~P.~R. Basis set
  convergence of the interaction energy of hydrogen-bonded complexes. \emph{J.
  Chem. Phys.} \textbf{1999}, \emph{111}, 9157--9167\relax
\mciteBstWouldAddEndPuncttrue
\mciteSetBstMidEndSepPunct{\mcitedefaultmidpunct}
{\mcitedefaultendpunct}{\mcitedefaultseppunct}\relax
\EndOfBibitem
\bibitem[Werner \latin{et~al.}(2007)Werner, Adler, and
  Manby]{werner2007general}
Werner,~H.-J.; Adler,~T.~B.; Manby,~F.~R. General orbital invariant MP2-F12
  theory. \emph{J. Chem. Phys.} \textbf{2007}, \emph{126}, 164102\relax
\mciteBstWouldAddEndPuncttrue
\mciteSetBstMidEndSepPunct{\mcitedefaultmidpunct}
{\mcitedefaultendpunct}{\mcitedefaultseppunct}\relax
\EndOfBibitem
\bibitem[Adler \latin{et~al.}(2007)Adler, Knizia, and Werner]{adler2007simple}
Adler,~T.~B.; Knizia,~G.; Werner,~H.-J. A simple and efficient CCSD(T)-F12
  approximation. \emph{J. Chem. Phys.} \textbf{2007}, \emph{127}, 221106\relax
\mciteBstWouldAddEndPuncttrue
\mciteSetBstMidEndSepPunct{\mcitedefaultmidpunct}
{\mcitedefaultendpunct}{\mcitedefaultseppunct}\relax
\EndOfBibitem
\bibitem[Knizia \latin{et~al.}(2009)Knizia, Adler, and
  Werner]{knizia2009simplified}
Knizia,~G.; Adler,~T.~B.; Werner,~H.-J. Simplified CCSD(T)-F12 methods: Theory
  and benchmarks. \emph{J. Chem. Phys.} \textbf{2009}, \emph{130}, 054104\relax
\mciteBstWouldAddEndPuncttrue
\mciteSetBstMidEndSepPunct{\mcitedefaultmidpunct}
{\mcitedefaultendpunct}{\mcitedefaultseppunct}\relax
\EndOfBibitem
\bibitem[Noga and Šimunek(2009)Noga, and Šimunek]{noga2009on}
Noga,~J.; Šimunek,~J. On the one-particle basis set relaxation in R12 based
  theories. \emph{Chem. Phys.} \textbf{2009}, \emph{356}, 1 -- 6\relax
\mciteBstWouldAddEndPuncttrue
\mciteSetBstMidEndSepPunct{\mcitedefaultmidpunct}
{\mcitedefaultendpunct}{\mcitedefaultseppunct}\relax
\EndOfBibitem
\bibitem[Weigend(2002)]{weigend2002fully}
Weigend,~F. A fully direct RI-HF algorithm: Implementation{,} optimised
  auxiliary basis sets{,} demonstration of accuracy and efficiency. \emph{Phys.
  Chem. Chem. Phys.} \textbf{2002}, \emph{4}, 4285--4291\relax
\mciteBstWouldAddEndPuncttrue
\mciteSetBstMidEndSepPunct{\mcitedefaultmidpunct}
{\mcitedefaultendpunct}{\mcitedefaultseppunct}\relax
\EndOfBibitem
\bibitem[Weigend \latin{et~al.}(2002)Weigend, Köhn, and
  Hättig]{weigend2002efficient}
Weigend,~F.; Köhn,~A.; Hättig,~C. Efficient use of the correlation consistent
  basis sets in resolution of the identity MP2 calculations. \emph{J. Chem.
  Phys.} \textbf{2002}, \emph{116}, 3175--3183\relax
\mciteBstWouldAddEndPuncttrue
\mciteSetBstMidEndSepPunct{\mcitedefaultmidpunct}
{\mcitedefaultendpunct}{\mcitedefaultseppunct}\relax
\EndOfBibitem
\bibitem[Weigend and Ahlrichs(2005)Weigend, and Ahlrichs]{weigend2005balanced}
Weigend,~F.; Ahlrichs,~R. {Balanced basis sets of split valence, triple zeta
  valence and quadruple zeta valence quality for H to Rn: Design and assessment
  of accuracy}. \emph{Phys. Chem. Chem. Phys.} \textbf{2005}, \emph{7},
  3297--3305\relax
\mciteBstWouldAddEndPuncttrue
\mciteSetBstMidEndSepPunct{\mcitedefaultmidpunct}
{\mcitedefaultendpunct}{\mcitedefaultseppunct}\relax
\EndOfBibitem
\bibitem[Yousaf and Peterson(2009)Yousaf, and Peterson]{yousaf2009optimized}
Yousaf,~K.~E.; Peterson,~K.~A. Optimized complementary auxiliary basis sets for
  explicitly correlated methods: aug-cc-pVnZ orbital basis sets. \emph{Chem.
  Phys. Lett.} \textbf{2009}, \emph{476}, 303 -- 307\relax
\mciteBstWouldAddEndPuncttrue
\mciteSetBstMidEndSepPunct{\mcitedefaultmidpunct}
{\mcitedefaultendpunct}{\mcitedefaultseppunct}\relax
\EndOfBibitem
\bibitem[Williams \latin{et~al.}(1995)Williams, Szalewicz, Moszynski, and
  Jeziorski]{williams1995dispersion}
Williams,~H.~L.; Szalewicz,~K.; Moszynski,~R.; Jeziorski,~B. Dispersion energy
  in the coupled pair approximation with noniterative inclusion of single and
  triple excitations. \emph{J. Chem. Phys.} \textbf{1995}, \emph{103},
  4586--4599\relax
\mciteBstWouldAddEndPuncttrue
\mciteSetBstMidEndSepPunct{\mcitedefaultmidpunct}
{\mcitedefaultendpunct}{\mcitedefaultseppunct}\relax
\EndOfBibitem
\bibitem[Parrish \latin{et~al.}(2013)Parrish, Hohenstein, and
  Sherrill]{parrish2013tractability}
Parrish,~R.~M.; Hohenstein,~E.~G.; Sherrill,~C.~D. Tractability gains in
  symmetry-adapted perturbation theory including coupled double excitations:
  CCD+ST(CCD) dispersion with natural orbital truncations. \emph{J. Chem.
  Phys.} \textbf{2013}, \emph{139}, 174102\relax
\mciteBstWouldAddEndPuncttrue
\mciteSetBstMidEndSepPunct{\mcitedefaultmidpunct}
{\mcitedefaultendpunct}{\mcitedefaultseppunct}\relax
\EndOfBibitem
\bibitem[Parrish \latin{et~al.}(2017)Parrish, Burns, Smith, Simmonett,
  DePrince, Hohenstein, Bozkaya, Sokolov, Di~Remigio, Richard, Gonthier, James,
  McAlexander, Kumar, Saitow, Wang, Pritchard, Verma, Schaefer, Patkowski,
  King, Valeev, Evangelista, Turney, Crawford, and Sherrill]{parrish2017psi4}
Parrish,~R.~M.; Burns,~L.~A.; Smith,~D. G.~A.; Simmonett,~A.~C.;
  DePrince,~A.~E.; Hohenstein,~E.~G.; Bozkaya,~U.; Sokolov,~A.~Y.;
  Di~Remigio,~R.; Richard,~R.~M.; Gonthier,~J.~F.; James,~A.~M.;
  McAlexander,~H.~R.; Kumar,~A.; Saitow,~M.; Wang,~X.; Pritchard,~B.~P.;
  Verma,~P.; Schaefer,~H.~F.; Patkowski,~K.; King,~R.~A.; Valeev,~E.~F.;
  Evangelista,~F.~A.; Turney,~J.~M.; Crawford,~T.~D.; Sherrill,~C.~D. {Psi4
  1.1: An Open-Source Electronic Structure Program Emphasizing Automation,
  Advanced Libraries, and Interoperability}. \emph{J. Chem. Theory Comput.}
  \textbf{2017}, \emph{13}, 3185--3197\relax
\mciteBstWouldAddEndPuncttrue
\mciteSetBstMidEndSepPunct{\mcitedefaultmidpunct}
{\mcitedefaultendpunct}{\mcitedefaultseppunct}\relax
\EndOfBibitem
\bibitem[Bart{\'o}k and Yates(2019)Bart{\'o}k, and
  Yates]{bartok2019regularized}
Bart{\'o}k,~A.~P.; Yates,~J.~R. {Regularized SCAN functional}. \emph{J. Chem.
  Phys.} \textbf{2019}, \emph{150}, 161101\relax
\mciteBstWouldAddEndPuncttrue
\mciteSetBstMidEndSepPunct{\mcitedefaultmidpunct}
{\mcitedefaultendpunct}{\mcitedefaultseppunct}\relax
\EndOfBibitem
\bibitem[Furness \latin{et~al.}(2020)Furness, Kaplan, Ning, Perdew, and
  Sun]{furness2020accurate}
Furness,~J.~W.; Kaplan,~A.~D.; Ning,~J.; Perdew,~J.~P.; Sun,~J. {Accurate and
  numerically efficient r2SCAN meta-generalized gradient approximation}.
  \emph{J. Phys. Chem. Lett.} \textbf{2020}, \emph{11}, 8208--8215\relax
\mciteBstWouldAddEndPuncttrue
\mciteSetBstMidEndSepPunct{\mcitedefaultmidpunct}
{\mcitedefaultendpunct}{\mcitedefaultseppunct}\relax
\EndOfBibitem
\bibitem[Mejía-Rodríguez and Trickey(2020)Mejía-Rodríguez, and
  Trickey]{mejia2020spin}
Mejía-Rodríguez,~D.; Trickey,~S.~B. {Spin-Crossover from a Well-Behaved,
  Low-Cost meta-GGA Density Functional}. \emph{J. Phys. Chem. A} \textbf{2020},
  \emph{124}, 9889--9894\relax
\mciteBstWouldAddEndPuncttrue
\mciteSetBstMidEndSepPunct{\mcitedefaultmidpunct}
{\mcitedefaultendpunct}{\mcitedefaultseppunct}\relax
\EndOfBibitem
\bibitem[Talman and Shadwick(1976)Talman, and Shadwick]{talman:1976:OEP}
Talman,~J.~D.; Shadwick,~W.~F. Optimized effective atomic central potential.
  \emph{Phys. Rev. A} \textbf{1976}, \emph{14}, 36--40\relax
\mciteBstWouldAddEndPuncttrue
\mciteSetBstMidEndSepPunct{\mcitedefaultmidpunct}
{\mcitedefaultendpunct}{\mcitedefaultseppunct}\relax
\EndOfBibitem
\bibitem[Bartlett \latin{et~al.}(2005)Bartlett, Grabowski, Hirata, and
  Ivanov]{bartlett2005exchange}
Bartlett,~R.~J.; Grabowski,~I.; Hirata,~S.; Ivanov,~S. The exchange-correlation
  potential in ab initio density functional theory. \emph{J. Chem. Phys.}
  \textbf{2005}, \emph{122}, 034104\relax
\mciteBstWouldAddEndPuncttrue
\mciteSetBstMidEndSepPunct{\mcitedefaultmidpunct}
{\mcitedefaultendpunct}{\mcitedefaultseppunct}\relax
\EndOfBibitem
\bibitem[{\'S}miga \latin{et~al.}(2020){\'S}miga, Marusiak, Grabowski, and
  Fabiano]{UHFOEP2sc}
{\'S}miga,~S.; Marusiak,~V.; Grabowski,~I.; Fabiano,~E. The ab initio density
  functional theory applied for spin-polarized calculations. \emph{J. Chem.
  Phys.} \textbf{2020}, \emph{152}, 054109\relax
\mciteBstWouldAddEndPuncttrue
\mciteSetBstMidEndSepPunct{\mcitedefaultmidpunct}
{\mcitedefaultendpunct}{\mcitedefaultseppunct}\relax
\EndOfBibitem
\bibitem[Stanton \latin{et~al.}(2007)Stanton, Gauss, Watts, Nooijen, Oliphant,
  Perera, Szalay, Lauderdale, Kucharski, Gwaltney, Beck, Balkov{\'{a}},
  Bernholdt, Baeck, Rozyczko, Sekino, Hober, and {R. J. Bartlett Integral
  packages included are VMOL (J. Alml\"{a}f and P.R. Taylor); VPROPS (P.
  Taylor) ABACUS; (T. Helgaker, H.J. Aa. Jensen, P. J{\"{o}}rgensen, J. Olsen,
  and P.R. Taylor)}]{acesII}
Stanton,~J.~F.; Gauss,~J.; Watts,~J.~D.; Nooijen,~M.; Oliphant,~N.;
  Perera,~S.~A.; Szalay,~P.; Lauderdale,~W.~J.; Kucharski,~S.; Gwaltney,~S.;
  Beck,~S.; Balkov{\'{a}},~A.; Bernholdt,~D.~E.; Baeck,~K.~K.; Rozyczko,~P.;
  Sekino,~H.; Hober,~C.; {R. J. Bartlett Integral packages included are VMOL
  (J. Alml\"{a}f and P.R. Taylor); VPROPS (P. Taylor) ABACUS; (T. Helgaker,
  H.J. Aa. Jensen, P. J{\"{o}}rgensen, J. Olsen, and P.R. Taylor)}, \emph{ACES
  II}; Quantum Theory Project: Gainesville, Florida, 2007\relax
\mciteBstWouldAddEndPuncttrue
\mciteSetBstMidEndSepPunct{\mcitedefaultmidpunct}
{\mcitedefaultendpunct}{\mcitedefaultseppunct}\relax
\EndOfBibitem
\bibitem[Śmiga and Fabiano(2017)Śmiga, and Fabiano]{Smi-gCC-2017}
Śmiga,~S.; Fabiano,~E. Approximate solution of coupled cluster equations:
  application to the coupled cluster doubles method and non-covalent
  interacting systems. \emph{Phys. Chem. Chem. Phys.} \textbf{2017}, \emph{19},
  30249--30260\relax
\mciteBstWouldAddEndPuncttrue
\mciteSetBstMidEndSepPunct{\mcitedefaultmidpunct}
{\mcitedefaultendpunct}{\mcitedefaultseppunct}\relax
\EndOfBibitem
\bibitem[Grabowski \latin{et~al.}(2014)Grabowski, Fabiano, Teale, {\'S}miga,
  Buksztel, and Sala]{OEPSOSszs}
Grabowski,~I.; Fabiano,~E.; Teale,~A.~M.; {\'S}miga,~S.; Buksztel,~A.;
  Sala,~F.~D. Orbital-dependent second-order scaled-opposite-spin correlation
  functionals in the optimized effective potential method. \emph{J. Chem.
  Phys.} \textbf{2014}, \emph{141}, 024113\relax
\mciteBstWouldAddEndPuncttrue
\mciteSetBstMidEndSepPunct{\mcitedefaultmidpunct}
{\mcitedefaultendpunct}{\mcitedefaultseppunct}\relax
\EndOfBibitem
\bibitem[Śmiga \latin{et~al.}(2019)Śmiga, Grabowski, Witkowski, Mussard, and
  Toulouse]{smiga2019self}
Śmiga,~S.; Grabowski,~I.; Witkowski,~M.; Mussard,~B.; Toulouse,~J.
  Self-consistent range-separated density-functional theory with second-order
  perturbative correction via the optimized-effective-potential method.
  \emph{J. Chem. Theory Comput.} \textbf{2019}, \emph{16}, 211--223\relax
\mciteBstWouldAddEndPuncttrue
\mciteSetBstMidEndSepPunct{\mcitedefaultmidpunct}
{\mcitedefaultendpunct}{\mcitedefaultseppunct}\relax
\EndOfBibitem
\bibitem[Śmiga and Constantin(2020)Śmiga, and Constantin]{OEPxPBE}
Śmiga,~S.; Constantin,~L.~A. Unveiling the Physics Behind Hybrid Functionals.
  \emph{J. Phys. Chem. A} \textbf{2020}, \emph{124}, 5606--5614\relax
\mciteBstWouldAddEndPuncttrue
\mciteSetBstMidEndSepPunct{\mcitedefaultmidpunct}
{\mcitedefaultendpunct}{\mcitedefaultseppunct}\relax
\EndOfBibitem
\bibitem[G{\"{o}}rling(1999)]{gorling:1999:OEP}
G{\"{o}}rling,~A. New {KS} method for molecules based on an exchange charge
  density generating the exact local {KS} exchange potential. \emph{Phys. Rev.
  Lett.} \textbf{1999}, \emph{83}, 5459--5462\relax
\mciteBstWouldAddEndPuncttrue
\mciteSetBstMidEndSepPunct{\mcitedefaultmidpunct}
{\mcitedefaultendpunct}{\mcitedefaultseppunct}\relax
\EndOfBibitem
\bibitem[Ivanov \latin{et~al.}(1999)Ivanov, Hirata, and
  Bartlett]{ivanov:1999:OEP}
Ivanov,~S.; Hirata,~S.; Bartlett,~R.~J. Exact exchange treatment for molecules
  in finite-basis-set {K}ohn-{S}ham theory. \emph{Phys. Rev. Lett.}
  \textbf{1999}, \emph{83}, 5455--5458\relax
\mciteBstWouldAddEndPuncttrue
\mciteSetBstMidEndSepPunct{\mcitedefaultmidpunct}
{\mcitedefaultendpunct}{\mcitedefaultseppunct}\relax
\EndOfBibitem
\bibitem[Hirata \latin{et~al.}(2001)Hirata, Ivanov, Grabowski, Bartlett, Burke,
  and Talman]{hirata:2001:OEPU}
Hirata,~S.; Ivanov,~S.; Grabowski,~I.; Bartlett,~R.~J.; Burke,~K.;
  Talman,~J.~D. Can optimized effective potentials be determined uniquely?
  \emph{J. Chem. Phys.} \textbf{2001}, \emph{115}, 1635--1649\relax
\mciteBstWouldAddEndPuncttrue
\mciteSetBstMidEndSepPunct{\mcitedefaultmidpunct}
{\mcitedefaultendpunct}{\mcitedefaultseppunct}\relax
\EndOfBibitem
\bibitem[Ivanov \latin{et~al.}(2002)Ivanov, Hirata, and
  Bartlett]{ivanov:2002:OEP}
Ivanov,~S.; Hirata,~S.; Bartlett,~R.~J. Finite-basis-set optimized effective
  potential exchange-only method. \emph{J. Chem. Phys.} \textbf{2002},
  \emph{116}, 1269--1276\relax
\mciteBstWouldAddEndPuncttrue
\mciteSetBstMidEndSepPunct{\mcitedefaultmidpunct}
{\mcitedefaultendpunct}{\mcitedefaultseppunct}\relax
\EndOfBibitem
\bibitem[Richard \latin{et~al.}(2014)Richard, Lao, and
  Herbert]{richard2013understanding}
Richard,~R.~M.; Lao,~K.~U.; Herbert,~J.~M. Understanding the many-body
  expansion for large systems. I. Precision considerations. \emph{J. Chem.
  Phys.} \textbf{2014}, \emph{141}, 014108\relax
\mciteBstWouldAddEndPuncttrue
\mciteSetBstMidEndSepPunct{\mcitedefaultmidpunct}
{\mcitedefaultendpunct}{\mcitedefaultseppunct}\relax
\EndOfBibitem
\bibitem[G{\'o}ra \latin{et~al.}(2011)G{\'o}ra, Podeszwa, Cencek, and
  Szalewicz]{gora2011interaction}
G{\'o}ra,~U.; Podeszwa,~R.; Cencek,~W.; Szalewicz,~K. Interaction energies of
  large clusters from many-body expansion. \emph{J. Chem. Phys.} \textbf{2011},
  \emph{135}, 224102\relax
\mciteBstWouldAddEndPuncttrue
\mciteSetBstMidEndSepPunct{\mcitedefaultmidpunct}
{\mcitedefaultendpunct}{\mcitedefaultseppunct}\relax
\EndOfBibitem
\bibitem[Marchetti and Werner(2009)Marchetti, and
  Werner]{marchetti2009accurate}
Marchetti,~O.; Werner,~H.-J. Accurate Calculations of Intermolecular
  Interaction Energies Using Explicitly Correlated Coupled Cluster Wave
  Functions and a Dispersion-Weighted MP2 Method. \emph{J. Phys. Chem. A}
  \textbf{2009}, \emph{113}, 11580--11585\relax
\mciteBstWouldAddEndPuncttrue
\mciteSetBstMidEndSepPunct{\mcitedefaultmidpunct}
{\mcitedefaultendpunct}{\mcitedefaultseppunct}\relax
\EndOfBibitem
\bibitem[Lao and Herbert(2018)Lao, and Herbert]{lao2018simple}
Lao,~K.~U.; Herbert,~J.~M. {A simple correction for nonadditive dispersion
  within extended symmetry-adapted perturbation theory (XSAPT)}. \emph{J. Chem.
  Theory Comput.} \textbf{2018}, \emph{14}, 5128--5142\relax
\mciteBstWouldAddEndPuncttrue
\mciteSetBstMidEndSepPunct{\mcitedefaultmidpunct}
{\mcitedefaultendpunct}{\mcitedefaultseppunct}\relax
\EndOfBibitem
\bibitem[Riplinger \latin{et~al.}(2013)Riplinger, Sandhoefer, Hansen, and
  Neese]{riplinger2013natural}
Riplinger,~C.; Sandhoefer,~B.; Hansen,~A.; Neese,~F. Natural triple excitations
  in local coupled cluster calculations with pair natural orbitals. \emph{J.
  Chem. Phys.} \textbf{2013}, \emph{139}, 134101\relax
\mciteBstWouldAddEndPuncttrue
\mciteSetBstMidEndSepPunct{\mcitedefaultmidpunct}
{\mcitedefaultendpunct}{\mcitedefaultseppunct}\relax
\EndOfBibitem
\bibitem[Liakos \latin{et~al.}(2015)Liakos, Sparta, Kesharwani, Martin, and
  Neese]{liakos2015exploring}
Liakos,~D.~G.; Sparta,~M.; Kesharwani,~M.~K.; Martin,~J.~M.; Neese,~F.
  {Exploring the Accuracy Limits of Local Pair Natural Orbital Coupled-Cluster
  Theory}. \emph{J. Chem. Theory Comput.} \textbf{2015}, \emph{11},
  1525--1539\relax
\mciteBstWouldAddEndPuncttrue
\mciteSetBstMidEndSepPunct{\mcitedefaultmidpunct}
{\mcitedefaultendpunct}{\mcitedefaultseppunct}\relax
\EndOfBibitem
\bibitem[Dubecký \latin{et~al.}(2019)Dubecký, Jurečka, Mitas, Ditte, and
  Fanta]{dubecky2019toward}
Dubecký,~M.; Jurečka,~P.; Mitas,~L.; Ditte,~M.; Fanta,~R. Toward Accurate
  Hydrogen Bonds by Scalable Quantum Monte Carlo. \emph{J. Chem. Theory
  Comput.} \textbf{2019}, \emph{15}, 3552--3557\relax
\mciteBstWouldAddEndPuncttrue
\mciteSetBstMidEndSepPunct{\mcitedefaultmidpunct}
{\mcitedefaultendpunct}{\mcitedefaultseppunct}\relax
\EndOfBibitem
\bibitem[Řezáč \latin{et~al.}(2015)Řezáč, Dubecký, Jurečka, and
  Hobza]{rezac2015extensions}
Řezáč,~J.; Dubecký,~M.; Jurečka,~P.; Hobza,~P. Extensions and applications
  of the A24 data set of accurate interaction energies. \emph{Phys. Chem. Chem.
  Phys.} \textbf{2015}, \emph{17}, 19268--19277\relax
\mciteBstWouldAddEndPuncttrue
\mciteSetBstMidEndSepPunct{\mcitedefaultmidpunct}
{\mcitedefaultendpunct}{\mcitedefaultseppunct}\relax
\EndOfBibitem
\bibitem[Dunning(1989)]{dunning1989gaussian}
Dunning,~T.~H.,~Jr. Gaussian basis sets for use in correlated molecular
  calculations. I. The atoms boron through neon and hydrogen. \emph{J. Chem.
  Phys.} \textbf{1989}, \emph{90}, 1007--1023\relax
\mciteBstWouldAddEndPuncttrue
\mciteSetBstMidEndSepPunct{\mcitedefaultmidpunct}
{\mcitedefaultendpunct}{\mcitedefaultseppunct}\relax
\EndOfBibitem
\bibitem[Peterson and Dunning(2002)Peterson, and Dunning]{peterson2002accurate}
Peterson,~K.~A.; Dunning,~T.~H. Accurate correlation consistent basis sets for
  molecular core–valence correlation effects: The second row atoms Al–Ar,
  and the first row atoms B–Ne revisited. \emph{J. Chem. Phys.}
  \textbf{2002}, \emph{117}, 10548--10560\relax
\mciteBstWouldAddEndPuncttrue
\mciteSetBstMidEndSepPunct{\mcitedefaultmidpunct}
{\mcitedefaultendpunct}{\mcitedefaultseppunct}\relax
\EndOfBibitem
\bibitem[Pitoňák \latin{et~al.}(2009)Pitoňák, Neogrády, Černý, Grimme,
  and Hobza]{pitonak2009scaled}
Pitoňák,~M.; Neogrády,~P.; Černý,~J.; Grimme,~S.; Hobza,~P. Scaled MP3
  Non-Covalent Interaction Energies Agree Closely with Accurate CCSD(T)
  Benchmark Data. \emph{ChemPhysChem} \textbf{2009}, \emph{10}, 282--289\relax
\mciteBstWouldAddEndPuncttrue
\mciteSetBstMidEndSepPunct{\mcitedefaultmidpunct}
{\mcitedefaultendpunct}{\mcitedefaultseppunct}\relax
\EndOfBibitem
\bibitem[Axilrod and Teller(1943)Axilrod, and
  Teller]{Axilrod-Teller:3Bdispersion}
Axilrod,~B.~M.; Teller,~E. Interaction of the van der Waals Type Between Three
  Atoms. \emph{J. Chem. Phys.} \textbf{1943}, \emph{11}, 299--300\relax
\mciteBstWouldAddEndPuncttrue
\mciteSetBstMidEndSepPunct{\mcitedefaultmidpunct}
{\mcitedefaultendpunct}{\mcitedefaultseppunct}\relax
\EndOfBibitem
\bibitem[Muto(1943)]{muto1943force}
Muto,~Y. Force between nonpolar molecules. \emph{J. Phys. Math. Soc. Japan}
  \textbf{1943}, \emph{17}, 629\relax
\mciteBstWouldAddEndPuncttrue
\mciteSetBstMidEndSepPunct{\mcitedefaultmidpunct}
{\mcitedefaultendpunct}{\mcitedefaultseppunct}\relax
\EndOfBibitem
\bibitem[kli()]{klimes2020git}
Outputs of calculations and processing script.
  \url{https://doi.org/10.5281/zenodo.4429677}\relax
\mciteBstWouldAddEndPuncttrue
\mciteSetBstMidEndSepPunct{\mcitedefaultmidpunct}
{\mcitedefaultendpunct}{\mcitedefaultseppunct}\relax
\EndOfBibitem
\bibitem[Valiron and Mayer(1997)Valiron, and Mayer]{valiron1997hierarchy}
Valiron,~P.; Mayer,~I. Hierarchy of counterpoise corrections for $N$-body
  clusters: generalization of the {Boys}-{Bernardi} scheme. \emph{Chem. Phys.
  Lett.} \textbf{1997}, \emph{275}, 46--55\relax
\mciteBstWouldAddEndPuncttrue
\mciteSetBstMidEndSepPunct{\mcitedefaultmidpunct}
{\mcitedefaultendpunct}{\mcitedefaultseppunct}\relax
\EndOfBibitem
\bibitem[Walczak \latin{et~al.}(2011)Walczak, Friedrich, and
  Dolg]{walczak2011on}
Walczak,~K.; Friedrich,~J.; Dolg,~M. On basis set superposition error corrected
  stabilization energies for large n-body clusters. \emph{J. Chem. Phys.}
  \textbf{2011}, \emph{135}\relax
\mciteBstWouldAddEndPuncttrue
\mciteSetBstMidEndSepPunct{\mcitedefaultmidpunct}
{\mcitedefaultendpunct}{\mcitedefaultseppunct}\relax
\EndOfBibitem
\bibitem[Richard \latin{et~al.}(2013)Richard, Lao, and
  Herbert]{richard2013achieving}
Richard,~R.; Lao,~K.; Herbert,~J. Achieving the CCSD(T) basis-set limit in
  sizable molecular clusters: Counterpoise corrections for the many-body
  expansion. \emph{J. Phys. Chem. Lett.} \textbf{2013}, \emph{4},
  2674--2680\relax
\mciteBstWouldAddEndPuncttrue
\mciteSetBstMidEndSepPunct{\mcitedefaultmidpunct}
{\mcitedefaultendpunct}{\mcitedefaultseppunct}\relax
\EndOfBibitem
\bibitem[Ouyang and Bettens(2015)Ouyang, and Bettens]{ouyang2015many}
Ouyang,~J.~F.; Bettens,~R. P.~A. Many-Body Basis Set Superposition Effect.
  \emph{J. Chem. Theory Comput.} \textbf{2015}, \emph{11}, 5132--5143\relax
\mciteBstWouldAddEndPuncttrue
\mciteSetBstMidEndSepPunct{\mcitedefaultmidpunct}
{\mcitedefaultendpunct}{\mcitedefaultseppunct}\relax
\EndOfBibitem
\bibitem[Liu and Herbert(2017)Liu, and Herbert]{liu2017understanding}
Liu,~K.-Y.; Herbert,~J.~M. Understanding the many-body expansion for large
  systems. III. Critical role of four-body terms, counterpoise corrections, and
  cutoffs. \emph{J. Chem. Phys.} \textbf{2017}, \emph{147}, 161729\relax
\mciteBstWouldAddEndPuncttrue
\mciteSetBstMidEndSepPunct{\mcitedefaultmidpunct}
{\mcitedefaultendpunct}{\mcitedefaultseppunct}\relax
\EndOfBibitem
\bibitem[Richard \latin{et~al.}(2018)Richard, Bakr, and
  Sherrill]{richard2018understanding}
Richard,~R.; Bakr,~B.; Sherrill,~C. Understanding the Many-Body Basis Set
  Superposition Error: Beyond Boys and Bernardi. \emph{J. Chem. Theory Comput.}
  \textbf{2018}, \emph{14}, 2386--2400\relax
\mciteBstWouldAddEndPuncttrue
\mciteSetBstMidEndSepPunct{\mcitedefaultmidpunct}
{\mcitedefaultendpunct}{\mcitedefaultseppunct}\relax
\EndOfBibitem
\bibitem[Peyton and Crawford(2019)Peyton, and Crawford]{peyton2019basis}
Peyton,~B.; Crawford,~T. Basis Set Superposition Errors in the Many-Body
  Expansion of Molecular Properties. \emph{J. Phys. Chem. A} \textbf{2019},
  \emph{123}, 4500--4511\relax
\mciteBstWouldAddEndPuncttrue
\mciteSetBstMidEndSepPunct{\mcitedefaultmidpunct}
{\mcitedefaultendpunct}{\mcitedefaultseppunct}\relax
\EndOfBibitem
\bibitem[Bartlett(2010)]{bartlett2010ab}
Bartlett,~R.~J. Ab initio DFT and its role in electronic structure theory.
  \emph{Mol. Phys.} \textbf{2010}, \emph{108}, 3299--3311\relax
\mciteBstWouldAddEndPuncttrue
\mciteSetBstMidEndSepPunct{\mcitedefaultmidpunct}
{\mcitedefaultendpunct}{\mcitedefaultseppunct}\relax
\EndOfBibitem
\bibitem[Yang \latin{et~al.}(2012)Yang, Cohen, and
  Mori-Sanchez]{yang2012derivative}
Yang,~W.; Cohen,~A.~J.; Mori-Sanchez,~P. Derivative discontinuity, bandgap and
  lowest unoccupied molecular orbital in density functional theory. \emph{J.
  Chem. Phys.} \textbf{2012}, \emph{136}, 204111\relax
\mciteBstWouldAddEndPuncttrue
\mciteSetBstMidEndSepPunct{\mcitedefaultmidpunct}
{\mcitedefaultendpunct}{\mcitedefaultseppunct}\relax
\EndOfBibitem
\bibitem[Lias \latin{et~al.}(2013)Lias, Bartmess, Liebman, Holmes, Levin, and
  Mallard]{linstrom2013nist}
Lias,~S.; Bartmess,~J.; Liebman,~J.; Holmes,~J.; Levin,~R.; Mallard,~W. In
  \emph{{NIST Chemistry WebBook, NIST Standard Reference Database Number 69}};
  Linstrom,~P., Mallard,~W., Eds.; National Institute of Standards and
  Technology, 2013; Chapter Ion Energetics Data\relax
\mciteBstWouldAddEndPuncttrue
\mciteSetBstMidEndSepPunct{\mcitedefaultmidpunct}
{\mcitedefaultendpunct}{\mcitedefaultseppunct}\relax
\EndOfBibitem
\bibitem[Van~Meer \latin{et~al.}(2014)Van~Meer, Gritsenko, and
  Baerends]{van2014physical}
Van~Meer,~R.; Gritsenko,~O.; Baerends,~E. {Physical meaning of virtual
  Kohn--Sham orbitals and orbital energies: an ideal basis for the description
  of molecular excitations}. \emph{J. Chem. Theory Comput.} \textbf{2014},
  \emph{10}, 4432--4441\relax
\mciteBstWouldAddEndPuncttrue
\mciteSetBstMidEndSepPunct{\mcitedefaultmidpunct}
{\mcitedefaultendpunct}{\mcitedefaultseppunct}\relax
\EndOfBibitem
\bibitem[Śmiga \latin{et~al.}(2016)Śmiga, Franck, Mussard, Buksztel,
  Grabowski, Luppi, and Toulouse]{smiga2016self}
Śmiga,~S.; Franck,~O.; Mussard,~B.; Buksztel,~A.; Grabowski,~I.; Luppi,~E.;
  Toulouse,~J. Self-consistent double-hybrid density-functional theory using
  the optimized-effective-potential method. \emph{J. Chem. Phys.}
  \textbf{2016}, \emph{145}, 144102\relax
\mciteBstWouldAddEndPuncttrue
\mciteSetBstMidEndSepPunct{\mcitedefaultmidpunct}
{\mcitedefaultendpunct}{\mcitedefaultseppunct}\relax
\EndOfBibitem
\bibitem[Chen \latin{et~al.}(2017)Chen, Ko, Remsing, Andrade, Santra, Sun,
  Selloni, Car, Klein, Perdew, and Wu]{chen2017ab}
Chen,~M.; Ko,~H.-Y.; Remsing,~R.~C.; Andrade,~M. F.~C.; Santra,~B.; Sun,~Z.;
  Selloni,~A.; Car,~R.; Klein,~M.~L.; Perdew,~J.~P.; Wu,~X. Ab initio theory
  and modeling of water. \emph{Proc. Natl. Acad. Sci. U. S. A.} \textbf{2017},
  \emph{114}, 10846--10851\relax
\mciteBstWouldAddEndPuncttrue
\mciteSetBstMidEndSepPunct{\mcitedefaultmidpunct}
{\mcitedefaultendpunct}{\mcitedefaultseppunct}\relax
\EndOfBibitem
\bibitem[Gillan \latin{et~al.}(2016)Gillan, Alfe, and
  Michaelides]{gillan2016perspective}
Gillan,~M.~J.; Alfe,~D.; Michaelides,~A. Perspective: How good is DFT for
  water? \emph{J. Chem. Phys.} \textbf{2016}, \emph{144}, 130901\relax
\mciteBstWouldAddEndPuncttrue
\mciteSetBstMidEndSepPunct{\mcitedefaultmidpunct}
{\mcitedefaultendpunct}{\mcitedefaultseppunct}\relax
\EndOfBibitem
\bibitem[Adamo \latin{et~al.}(1999)Adamo, Cossi, Scalmani, and
  Barone]{adamo1999accurate}
Adamo,~C.; Cossi,~M.; Scalmani,~G.; Barone,~V. {Accurate static
  polarizabilities by density functional theory: assessment of the PBE0 model}.
  \emph{Chem. Phys. Lett.} \textbf{1999}, \emph{307}, 265--271\relax
\mciteBstWouldAddEndPuncttrue
\mciteSetBstMidEndSepPunct{\mcitedefaultmidpunct}
{\mcitedefaultendpunct}{\mcitedefaultseppunct}\relax
\EndOfBibitem
\bibitem[Wen \latin{et~al.}(2012)Wen, Nanda, Huang, and
  Beran]{BeranPCCP2012:fragment}
Wen,~S.; Nanda,~K.; Huang,~Y.; Beran,~G. J.~O. Practical quantum
  mechanics-based fragment methods for predicting molecular crystal properties.
  \emph{Phys. Chem. Chem. Phys.} \textbf{2012}, \emph{14}, 7578--7590\relax
\mciteBstWouldAddEndPuncttrue
\mciteSetBstMidEndSepPunct{\mcitedefaultmidpunct}
{\mcitedefaultendpunct}{\mcitedefaultseppunct}\relax
\EndOfBibitem
\bibitem[Hermann and Schwerdtfeger(2008)Hermann, and
  Schwerdtfeger]{hermann2008ground}
Hermann,~A.; Schwerdtfeger,~P. Ground-State Properties of Crystalline Ice from
  Periodic Hartree-Fock Calculations and a Coupled-Cluster-Based Many-Body
  Decomposition of the Correlation Energy. \emph{Phys. Rev. Lett.}
  \textbf{2008}, \emph{101}, 183005\relax
\mciteBstWouldAddEndPuncttrue
\mciteSetBstMidEndSepPunct{\mcitedefaultmidpunct}
{\mcitedefaultendpunct}{\mcitedefaultseppunct}\relax
\EndOfBibitem
\bibitem[Liu and Herbert(2020)Liu, and Herbert]{liu2020energy}
Liu,~K.-Y.; Herbert,~J.~M. Energy-Screened Many-Body Expansion: A Practical Yet
  Accurate Fragmentation Method for Quantum Chemistry. \emph{J. Chem. Theory
  Comput.} \textbf{2020}, \emph{16}, 475--487\relax
\mciteBstWouldAddEndPuncttrue
\mciteSetBstMidEndSepPunct{\mcitedefaultmidpunct}
{\mcitedefaultendpunct}{\mcitedefaultseppunct}\relax
\EndOfBibitem
\bibitem[Červinka and Beran(2018)Červinka, and Beran]{cervinka2018abinitio}
Červinka,~C.; Beran,~G. J.~O. Ab initio prediction of the polymorph phase
  diagram for crystalline methanol. \emph{Chem. Sci.} \textbf{2018}, \emph{9},
  4622--4629\relax
\mciteBstWouldAddEndPuncttrue
\mciteSetBstMidEndSepPunct{\mcitedefaultmidpunct}
{\mcitedefaultendpunct}{\mcitedefaultseppunct}\relax
\EndOfBibitem
\bibitem[Tuma and Sauer(2004)Tuma, and Sauer]{tuma2004hybrid}
Tuma,~C.; Sauer,~J. A hybrid MP2/planewave-DFT scheme for large chemical
  systems: proton jumps in zeolites. \emph{Chem. Phys. Lett.} \textbf{2004},
  \emph{387}, 388\relax
\mciteBstWouldAddEndPuncttrue
\mciteSetBstMidEndSepPunct{\mcitedefaultmidpunct}
{\mcitedefaultendpunct}{\mcitedefaultseppunct}\relax
\EndOfBibitem
\bibitem[Bludsk\'y \latin{et~al.}(2008)Bludsk\'y, Rube\v{s}, Sold\'an, and
  Nachtigal]{bludsky2008investigation}
Bludsk\'y,~O.; Rube\v{s},~M.; Sold\'an,~P.; Nachtigal,~P. Investigation of the
  benzene-dimer potential energy surface: DFT/CCSD(T) correction scheme.
  \emph{J. Chem. Phys.} \textbf{2008}, \emph{128}, {114102}\relax
\mciteBstWouldAddEndPuncttrue
\mciteSetBstMidEndSepPunct{\mcitedefaultmidpunct}
{\mcitedefaultendpunct}{\mcitedefaultseppunct}\relax
\EndOfBibitem
\bibitem[Liao and Grüneis(2016)Liao, and Grüneis]{liao2016finite}
Liao,~K.; Grüneis,~A. Communication: Finite size correction in periodic
  coupled cluster theory calculations of solids. \emph{J. Chem. Phys.}
  \textbf{2016}, \emph{145}, 141102\relax
\mciteBstWouldAddEndPuncttrue
\mciteSetBstMidEndSepPunct{\mcitedefaultmidpunct}
{\mcitedefaultendpunct}{\mcitedefaultseppunct}\relax
\EndOfBibitem
\end{mcitethebibliography}


\end{document}